\documentclass[aps,prd,floatfix,preprintnumbers,twocolumn,groupedaddress,nofootinbib,longbibliography,10pt]{revtex4-1}
\usepackage{mathtools,amssymb,braket,mathrsfs,tikz,xspace,xcolor,float,color,siunitx,comment,afterpage,multirow,array,hhline,tensor,hyperref, textcomp,tikz}
\usepackage[utf8]{inputenc}

\definecolor{patchcolor}{rgb}{1,1,1}

\providecommand{\href}[2]{#2}
\renewcommand{\O}{\mathcal{O}}

\newcommand{\I}{\mathcal{I}}
\DeclareMathOperator{\tr}{tr}

\DeclareRobustCommand{\Sec}[1]{Sec.~\ref{#1}}
\DeclareRobustCommand{\Secs}[2]{Secs.~\ref{#1} and \ref{#2}}

\DeclareRobustCommand{\App}[1]{App.~\ref{#1}}
\DeclareRobustCommand{\Tab}[1]{Table~\ref{#1}}

\DeclareRobustCommand{\Fig}[1]{Fig.~\ref{#1}}

\DeclareRobustCommand{\Eq}[1]{Eq.~(\ref{#1})}
\DeclareRobustCommand{\Eqs}[2]{Eqs.~(\ref{#1}) and (\ref{#2})}

\DeclareRobustCommand{\Ref}[1]{Ref.~\cite{#1}}

\DeclareRobustCommand{\dotgraph}[1]{\begin{gathered}\includegraphics[scale=#1]{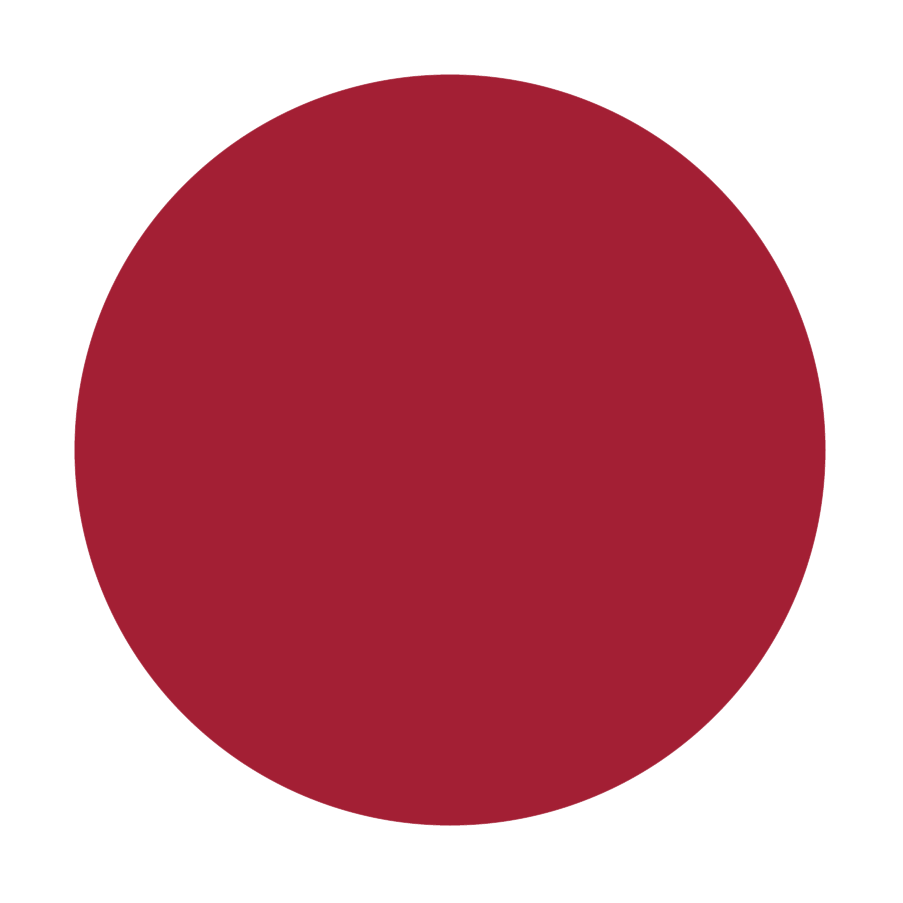}\end{gathered}}
\DeclareRobustCommand{\edgegraph}[1]{\begin{gathered}\includegraphics[scale=#1]{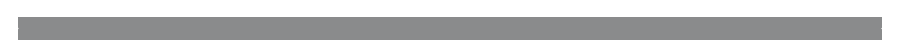}\end{gathered}}
\DeclareRobustCommand{\linegraph}[1]{\begin{gathered}\includegraphics[scale=#1]{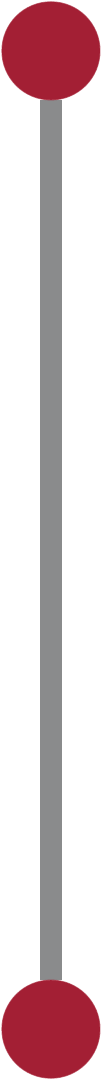}\end{gathered}}
\DeclareRobustCommand{\dumbbell}[1]{\begin{gathered}\includegraphics[scale=#1]{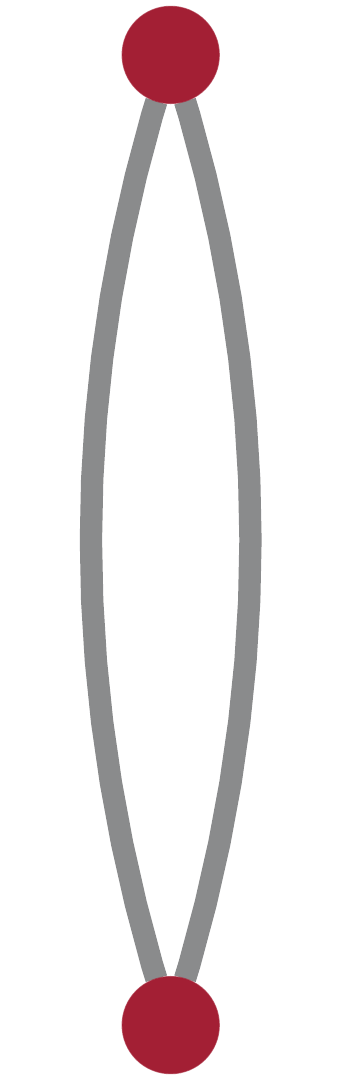}\end{gathered}}
\DeclareRobustCommand{\tribell}[1]{\begin{gathered}\includegraphics[scale=#1]{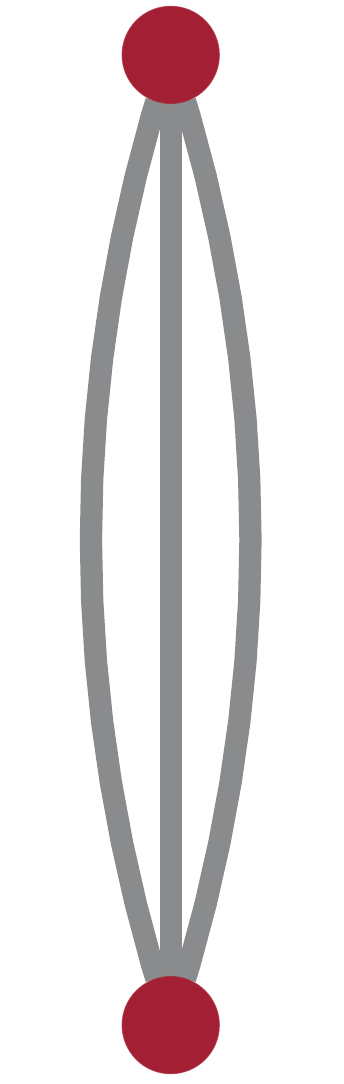}\end{gathered}}
\DeclareRobustCommand{\wedge}[1]{\begin{gathered}\includegraphics[scale=#1]{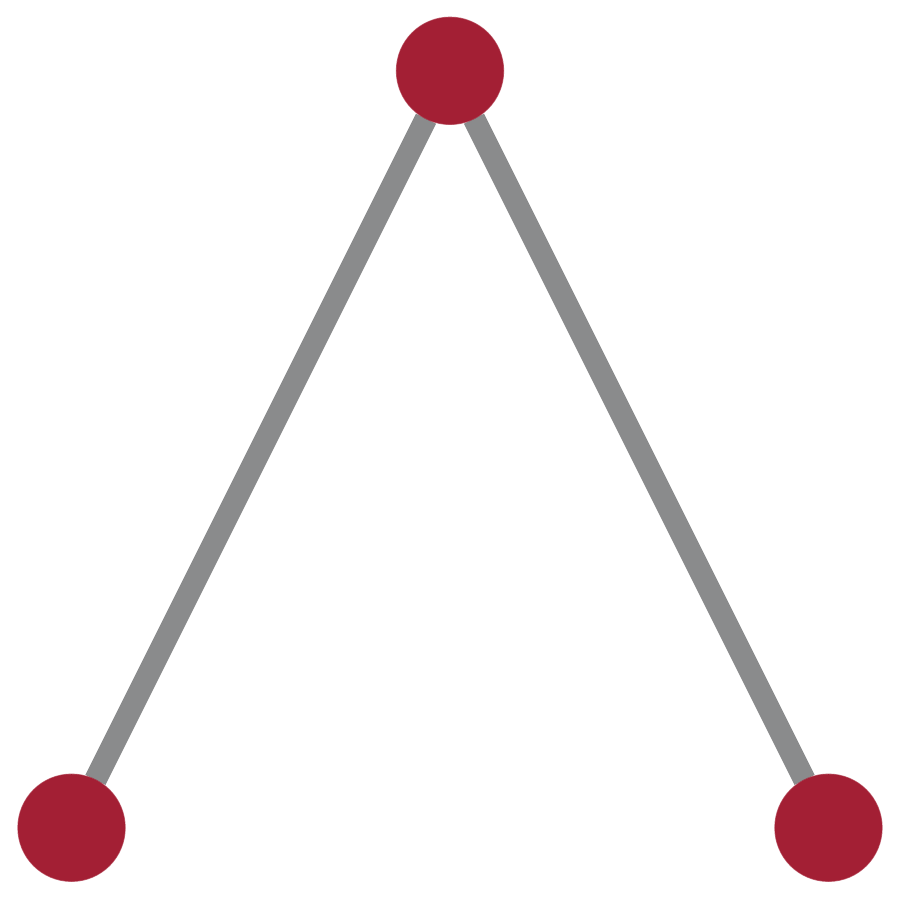}\end{gathered}}
\DeclareRobustCommand{\triangle}[1]{\begin{gathered}\includegraphics[scale=#1]{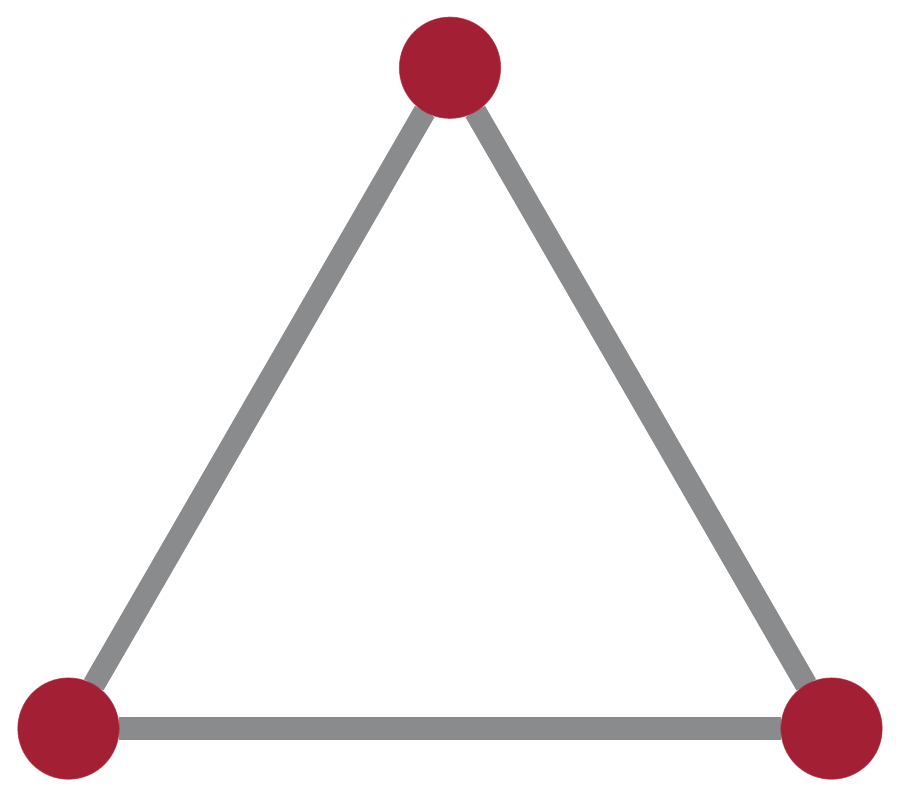}\end{gathered}}

\DeclareRobustCommand{\kite}[1]{\begin{gathered}\includegraphics[scale=#1]{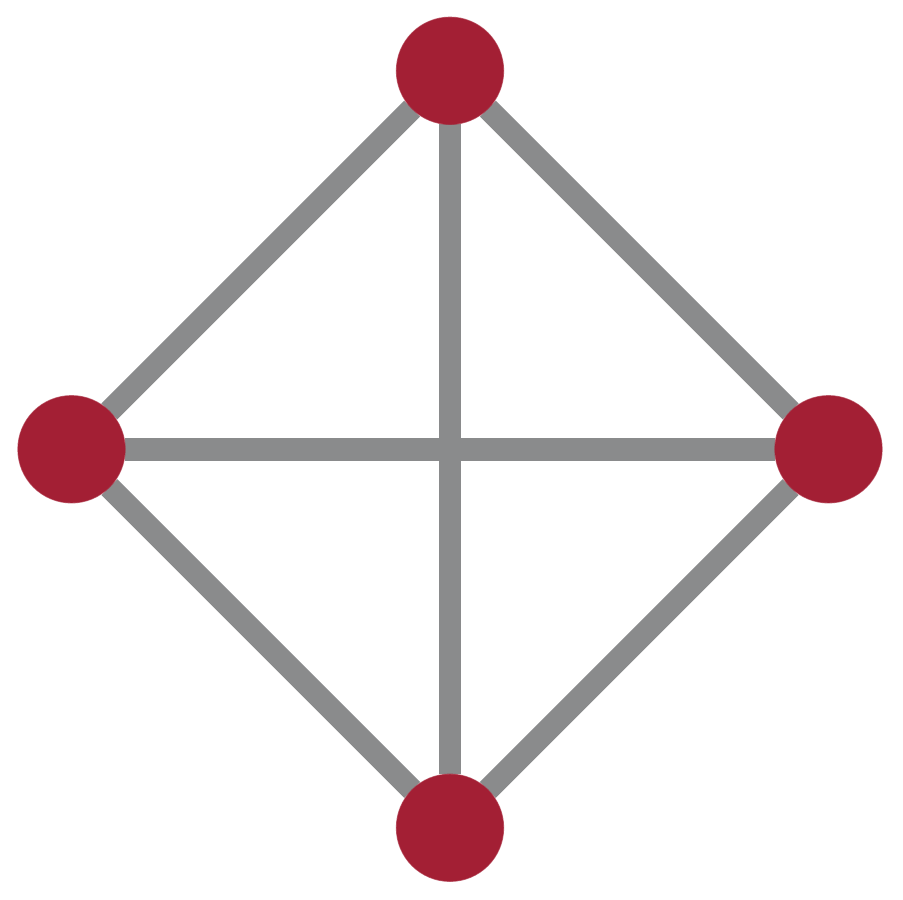}\end{gathered}}
\DeclareRobustCommand{\asymwedge}[1]{\begin{gathered}\includegraphics[scale=#1]{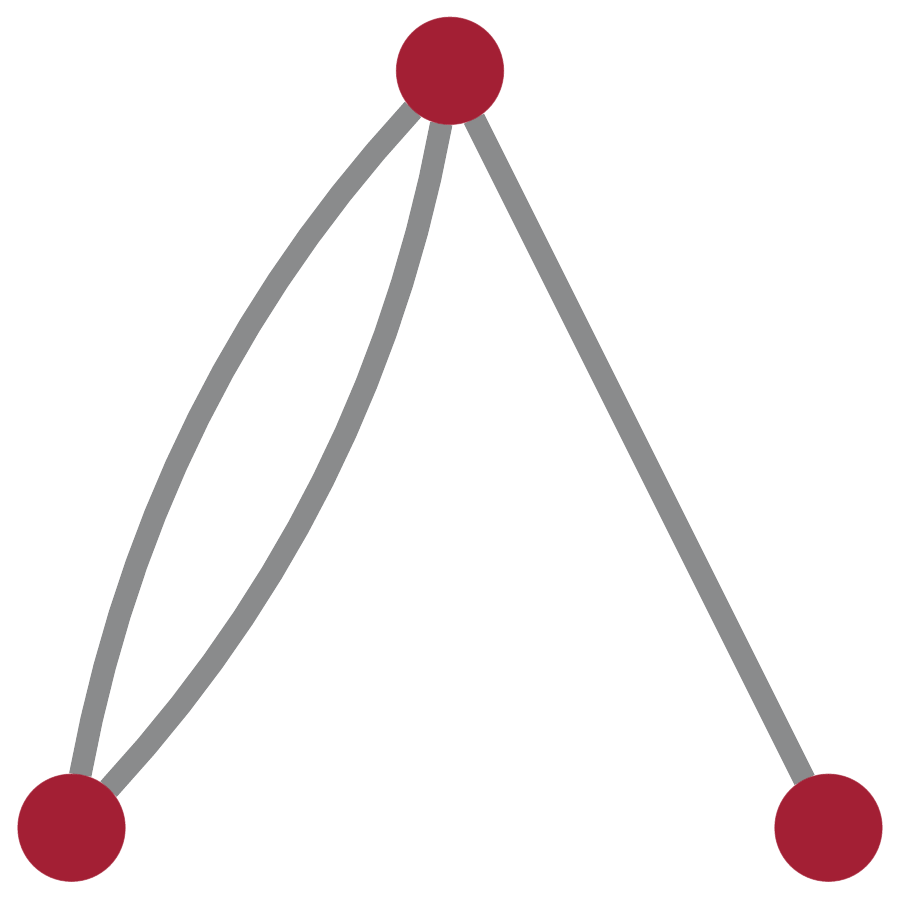}\end{gathered}}
\DeclareRobustCommand{\asymwedgeA}[1]{\begin{gathered}\includegraphics[scale=#1]{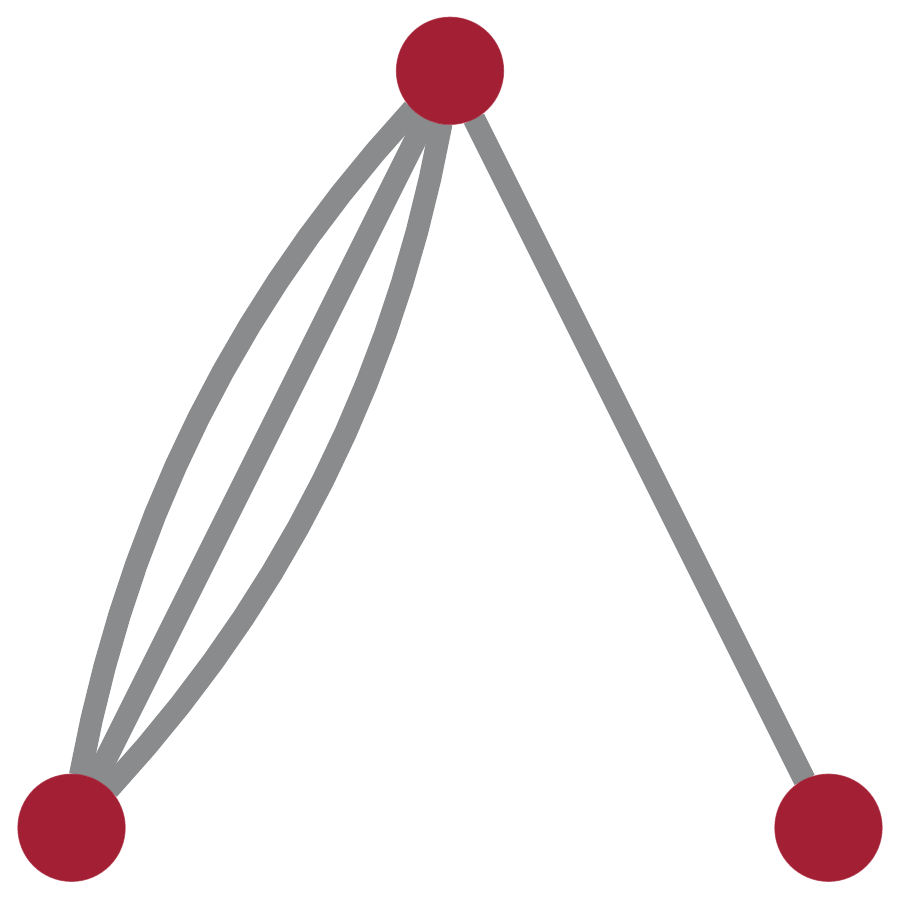}\end{gathered}}

\definecolor{jdtcolor}{rgb}{0.8,0,0}
\definecolor{emmcolor}{rgb}{0,0.8,0}
\definecolor{ptkcolor}{rgb}{0,0,0.8}

\begin{document}

\title{Cutting Multiparticle Correlators Down to Size}
\preprint{MIT-CTP 5150}

\author{Patrick T. Komiske}
\email{pkomiske@mit.edu}

\author{Eric M. Metodiev}
\email{metodiev@mit.edu}

\author{Jesse Thaler}
\email{jthaler@mit.edu}

\affiliation{Center for Theoretical Physics, Massachusetts Institute of Technology, Cambridge, MA 02139, USA}
\affiliation{Department of Physics, Harvard University, Cambridge, MA 02138, USA}

\begin{abstract}
Multiparticle correlators are mathematical objects frequently encountered in quantum field theory and collider physics.
By translating multiparticle correlators into the language of graph theory, we can gain new insights into their structure as well as identify efficient ways to manipulate them.
In this paper, we highlight the power of this graph-theoretic approach by ``cutting open'' the vertices and edges of the graphs, allowing us to systematically classify linear relations among multiparticle correlators and develop faster methods for their computation.
The naive computational complexity of an $N$-point correlator among $M$ particles is $\O(M^N)$, but when the pairwise distances between particles can be cast as an inner product, we show that all such correlators can be computed in linear $\O(M)$ runtime.
With the help of new tensorial objects called Energy Flow Moments, we achieve a fast implementation of jet substructure observables like $C_2$ and $D_2$, which are widely used at the Large Hadron Collider to identify boosted hadronic resonances.
As another application, we compute the number of leafless multigraphs with $d$ edges up to $d = 16$ (15,641,159), conjecturing that this is the same as the number of independent kinematic polynomials of degree $d$, previously known only to $d=8$ (279).
\end{abstract}

\flushbottom

\maketitle

{\small\tableofcontents}
\newpage

\section{Introduction}
\label{sec:intro}

Multiparticle correlators are ubiquitous mathematical structures that appear in a variety of physics domains, including kinematic polynomials for scattering amplitudes \cite{Boels:2013jua,oeisA226919}, operator bases for effective field theories \cite{Hogervorst:2014rta,Henning:2017fpj}, and many-body expansions for molecular analyses~\cite{PhysRevB.46.2133,doi:10.1021/ct9004917}.
Broadly speaking, multiparticle correlators appear whenever the fundamental entities of a system are arranged in sets: unordered, variable-length collections of objects.
These set elements may be nuclei, atoms, fields, particles, or other objects.
Here, we refer to them as ``particles'', since we will eventually focus on multiparticle correlators in collider physics~\cite{Larkoski:2013eya,Larkoski:2014gra,Komiske:2017aww}.

The basic structure of an $N$-point multiparticle correlator of $M$ particles is:
\begin{equation}
\label{eq:multcorr}
\sum_{i_1=1}^M\sum_{i_2=1}^M\cdots\sum_{i_N=1}^M z_{i_1}z_{i_2}\cdots z_{i_N} f(p_{i_1}, p_{i_2},\cdots, p_{i_N}),
\end{equation}
where $z_i$ are particle weights, $p_i$ are particle properties, and $f$ is a function of these properties.
In the particle physics literature, \Eq{eq:multcorr} appears as the $C$-correlators of~\Ref{Tkachov:1995kk}.
The function $f$ is often decomposed in terms of monomials of pairwise distances or invariants $\theta_{ij}$ between the particles:
\begin{equation}
\label{eq:fdecomp}
f(p_{i_1},\cdots,p_{i_N}) =  \theta_{i_1i_2}^{\alpha_{12}} \theta_{i_1 i_3}^{\alpha_{13}}\cdots \theta_{i_{N-1}i_N}^{\alpha_{N-1\,N}},
\end{equation}
where $\alpha_{ij}$ are integer exponents.
An argument for the generality of this restriction (up to isometries) in the context of collider observables is given in \Ref{Komiske:2017aww}.
This simplification allows multiparticle correlators to be decomposed into smaller pieces, which is essential for deriving the results below.

In this paper, we develop the theory of multiparticle correlators by representing them as multigraphs, allowing us to obtain many useful results by manipulating their vertices and edges.
Specifically, by ``cutting'' the multigraphs along either their vertices or edges, we derive otherwise opaque linear relations among sets of correlators.
Furthermore, in special cases where the pairwise distances $\theta_{ij}$ take the form of an inner product, we show that the complexity of computing a multiparticle correlator can be reduced from $\O(M^N)$ down to the minimally required $\O(M)$.
We highlight various connections of these mathematical results to high-energy physics, including counting kinematic polynomials relevant for superstring amplitudes~\cite{Boels:2013jua,oeisA226919} and speeding up the computation of jet substructure observables~\cite{Larkoski:2013eya,Larkoski:2014gra,Komiske:2017aww}.

A detailed outline of this paper is as follows.
In \Sec{sec:multcorr}, we summarize the basics of multiparticle correlators and review the multigraph-multiparticle correlator correspondence of \Ref{Komiske:2017aww}.
Several well-known examples of multiparticle correlators are discussed, including kinematic polynomials and Energy Flow Polynomials (EFPs)~\cite{Komiske:2017aww}.
We also review the computational complexity of multiparticle correlators and show how the simplification in \Eq{eq:fdecomp} allows us to improve upon the naive $\O(M^N)$ scaling using the variable elimination algorithm~\cite{Komiske:2017aww, zhang1996exploiting}.

In \Sec{sec:verts}, we introduce a method to slice the vertices of multigraphs.
By cutting vertices in half, the multiparticle correlator in \Eq{eq:multcorr} can be decomposed into contractions of ``particle tensors'' whose indices run from 1 to $M$ (the number of particles).
We then derive linear relations among multiparticle correlators when $M$ is small compared to $N$.
The key for understanding these linear relations is a well-known tensor antisymmetrization identity.
In $m$ dimensions, antisymmetrizing any tensor over $\ell>m$ indices yields zero:
\begin{equation}
\label{eq:antisymid}
T^{a_1\cdots a_j}_{b_1\cdots b_k [c_1 \cdots c_\ell]} = 0.
\end{equation}
This follows immediately from the fact that any assignment of the $m$ possible values to the $\ell > m$ indices must have a repetition and therefore vanish.
These particle tensors also enable further computational speedups through methods such as fast matrix multiplication and dynamic programming.

In \Sec{sec:edges}, we explore the consequences of making an additional simplifying assumption, namely that the pairwise distances $\theta_{ij}$ are given by an inner product:
\begin{equation}
\label{eq:innerprod}
\theta_{ij} = \eta_{\mu\nu} u_i^\mu u_j^\nu,
\end{equation}
for some vectors $u_i^\mu,u_j^\nu$ and metric $\eta_{\mu\nu}$ with Einstein summation convention.
This assumption applies, for instance, to kinematic polynomials of Mandelstam invariants and EFPs with angular exponent $\beta=2$.
Having made this assumption, multigraph edges can be sliced and the multiparticle correlators can be expressed as contractions of 
``moment tensors''.
This allows us to derive new linear relations and computational speedups by cutting edges of the multigraphs.
The antisymmetrization identity in \Eq{eq:antisymid}, now applied to the moment tensors, explains redundancies that appear when the dimension of the inner product space is small compared to the number of edges.
Multiparticle correlators can now be computed in $\O(M)$, since they are sewn together from moment tensors, each of which is $\O(M)$ to compute.
This result generalizes the well-known fact that the invariant mass of a set of massless particles does not require $\O(M^2)$ time to compute, since one can simply sum over the four-vectors in $\O(M)$:
\begin{equation}
\label{eq:massexample}
\sum_{i = 1}^M \sum_{j = 1}^M p_i \cdot p_j = \left(\sum_{i = 1}^{M} p_i^\mu \right)^2.
\end{equation}
This reduced complexity is particularly relevant for quantum algorithms for collider physics, where preprocessing and loading classical data into a quantum computer can be a key computational bottleneck~\cite{Wei:2019rqy}.

In \Sec{sec:leaves}, we show how momentum conservation can be used to trim away valency-one leaves from the multigraphs.
As an interesting physical application, we use these trimmed moment tensors to count independent kinematic polynomials up to degree $d$, which is relevant for enumerating superstring amplitudes~\cite{Boels:2013jua}.
By translating this problem to our graphical language, we are able to significantly extend existing integer sequences by simply enumerating leafless multigraphs, advancing previous results from $d = 8$~\cite{Boels:2013jua,oeisA226919} to $d = 16$~\cite{oeisA307317,oeisA307316}.

In \Sec{sec:energyflow}, we highlight the relevance of our results for collider physics.
As shown in \Ref{Komiske:2017aww}, the EFPs fully capture the infrared-and-collinear-safe (IRC-safe) information in the radiation pattern of an event, for any choice of angular exponent $\beta$.
For the special choice of $\beta = 2$, the EFPs can be written as Lorentz contractions of Energy Flow Moments (EFMs), which are IRC safe by construction.
An EFM with $v$ indices takes the form:
\begin{equation}
\label{eq:efm}
\mathcal I^{\mu_1 \cdots \mu_v} = 2^{v/2} \sum_{i=1}^M E_i \, n_i^{\mu_1} \cdots n_i^{\mu_v},
\end{equation}
where $E_i$ are particle energies, $p_i^\mu$ are massless particle momenta, and $n_i^\mu \equiv p_i^\mu/E_i$.
The EFMs are introduced for the first time in this paper, though we show that they are closely related to other moment tensors appearing previously in the literature.
Using the EFMs, we demonstrate that the computationally expensive jet substructure observables $C_2$~\cite{Larkoski:2013eya} and $D_2$~\cite{Larkoski:2014gra}, widely applied for new physics searches at the Large Hadron Collider (LHC), can significantly benefit from the computational speedups developed here.
We make the fast implementations of these observables available in the \href{https://energyflow.network}{\textsc{EnergyFlow}} Python package~\cite{EnergyFlow}, which also provides tools to compute any multiparticle correlator.

Our conclusions are presented in \Sec{sec:conc}.
We derive additional linear relations for multiparticle correlators in the context of $e^+e^-$ collisions in \App{app:euclideanslicing}.

\clearpage

\section{Multiparticle Correlators}
\label{sec:multcorr}

This section summarizes the basic properties of multiparticle correlators.
We review the correspondence between multiparticle correlators and multigraphs, provide two key examples of multiparticle correlators relevant for high-energy physics, and discuss previously known techniques to decrease their computational complexity.

\subsection{Correlators as Graphs}
\label{sec:corrgraphs}

As shown in \Ref{Komiske:2017aww}, multigraphs are an efficient and intuitive way to represent multiparticle correlators.
Each of the $N$ sums and weight factors in \Eq{eq:multcorr} corresponds to a vertex, and each pairwise distance factor in \Eq{eq:fdecomp} corresponds to an edge.
These rules can be graphically summarized as:
\begin{equation}
\label{eq:graphrules}
\dotgraph{0.04}_i = \sum_{i=1}^M z_{i}, \quad \quad \,_j\edgegraph{0.2}_k = \theta_{jk}.
\end{equation}
The ``multi'' in multigraph refers to the fact that two vertices can be connected by more than one edge.

In this language, a multiparticle correlator with graph $G$ is written as:
\begin{equation}
\label{eq:multcorrG}
\sum_{i_1=1}^M\sum_{i_2=1}^M\cdots\sum_{i_N=1}^Mz_{i_1}\cdots z_{i_N}\prod_{(k,\ell)\in G}\theta_{i_k i_\ell},
\end{equation}
where $(k,\ell)$ are the pairs of vertices connected by the edges in $G$.
As an example, consider the following multigraph and its corresponding correlator:
\begin{equation}
\asymwedge{0.16} = \sum_{i_1=1}^M \sum_{i_2=1}^M\sum_{i_3=1}^M z_{i_1}z_{i_2}z_{i_3} \theta_{i_1i_2}^2 \theta_{i_1i_3}.
\end{equation}
This correlator has three sums corresponding to the three vertices of the multigraph.
Each of the edges contributes a factor of $\theta_{ij}$ to the argument, with $\theta_{i_2i_3}$ absent due to no edge connecting two of the vertices.
Any permutation $\sigma$ of $i_1,\cdots,i_N$ to $i_{\sigma(1)},\cdots,i_{\sigma(N)}$ yields an identical correlator by the symmetry of the sum structure.
Graphically, this translates to the fact that isomorphic multigraphs correspond to the same correlator, which allows us to write the graphs without labels on the vertices.

To quantify the number of unique analytic structures in the correlators, it is helpful to organize them by the degree $d$ of the monomial, or equivalently the number of edges in the associated multigraph.
The number of unique multiparticle correlators of each $d$ up to $d=7$ is tabulated in \Tab{tab:oeisnumgraphs} based on entries in the On-Line Encyclopedia of Integer Sequences (OEIS)~\cite{oeis}.
As an explicit demonstration, \Tab{tab:corrlist} shows the expressions and multigraphs for all 13 multiparticle correlators with degree $d\le 3$.
Correlators corresponding to connected multigraphs are also separately tabulated, as disconnected ones are simply the product of their connected components.

\begin{table}[t]
\begin{tabular}{rr|rrrrrrrr}
\hline\hline
 & \multicolumn{1}{r|}{Degree $d$\quad\quad} &\, 0 & 1 & 2 & 3 & 4 & 5  & 6 & 7 \\ \hline \hline
\multirow{2}{18mm}{Connected}  & \href{https://oeis.org/A076864}{A076864}~\cite{oeisA076864} \,&\, 1 & 1 & 2 & 5 & 12 & 33 & 103 & 333  \\ 
 & Cumulative \,&\, 1 & 2 & 4 & 9 & 21 & 54 & 157 & 490 \\  \hline
\multirow{2}{18mm}{All} & \href{https://oeis.org/A050535}{A050535}~\cite{oeisA050535} \,&\, 1 & 1 & 3 & 8 & 23 & 66 & 212 & 686 \\
 & Cumulative \,&\, 1 & 2 & 5 & 13 & 36 & 102 & 314 & 1\,000 \\
 \hline\hline
\end{tabular}
\caption{\label{tab:oeisnumgraphs}
The number of multiparticle correlators of degree $d$.
These are equivalent to the number of non-isomorphic multigraphs with $d$ edges, tabulated for connected and all multigraphs.}
\end{table}

\begin{table*}[!p]
\begin{tabular}{@{$\quad$} c @{$\quad$} | @{$\quad$} c c @{$\quad$} l @{$\quad$}}
\hline
\hline
Degree & Multigraph & & Multiparticle Correlator \\
\hline
\hline
$d=0$ & $\dotgraph{0.02}$ & = & $\displaystyle\sum_{i=1}^M z_i$\\ \hline
& & & \\
$d=1$ & $\linegraph{0.12}$ & = & $\displaystyle\sum_{i_1=1}^M \sum_{i_2=1}^M z_{i_1} z_{i_2} \theta_{i_1i_2}$ \\
& & & \\ \hline
& & & \\
& $\dumbbell{0.12}$ & = & $\displaystyle\sum_{i_1=1}^M \sum_{i_2=1}^M z_{i_1} z_{i_2} \theta_{i_1i_2}^2 $ \\
$d=2$ & $\wedge{0.12}$ & = & $\displaystyle\sum_{i_1=1}^M \sum_{i_2=1}^M \sum_{i_3=1}^M z_{i_1} z_{i_2} z_{i_3} \theta_{i_1i_2} \theta_{i_1i_3}$ \\
 & $\linegraph{0.12}\,\linegraph{0.12}$ & = & $\displaystyle\sum_{i_1=1}^M \sum_{i_2=1}^M \sum_{i_3=1}^M \sum_{i_4=1}^M z_{i_1} z_{i_2} z_{i_3} z_{i_4} \theta_{i_1i_2} \theta_{i_3i_4}$ \\
 & & & \\ \hline
 & & & \\
& $\tribell{0.12}$ & = & $\displaystyle\sum_{i_1=1}^M \sum_{i_2=1}^M z_{i_1} z_{i_2} \theta_{i_1i_2}^3 $ \\
& $\triangle{0.12}$ & = & $\displaystyle\sum_{i_1=1}^M \sum_{i_2=1}^M \sum_{i_3=1}^M z_{i_1} z_{i_2} z_{i_3} \theta_{i_1i_2}\theta_{i_2i_3}\theta_{i_1i_3} $ \\
& $\begin{gathered}\includegraphics[scale=0.12]{graphs/3_3_2.pdf}\end{gathered}$&=&
$\displaystyle\sum_{i_1=1}^M\sum_{i_2=1}^M\sum_{i_3=1}^M z_{i_1}z_{i_2}z_{i_3}\theta_{i_1i_2}^2 \theta_{i_1i_3}$\\
& $\begin{gathered}\includegraphics[scale=0.14]{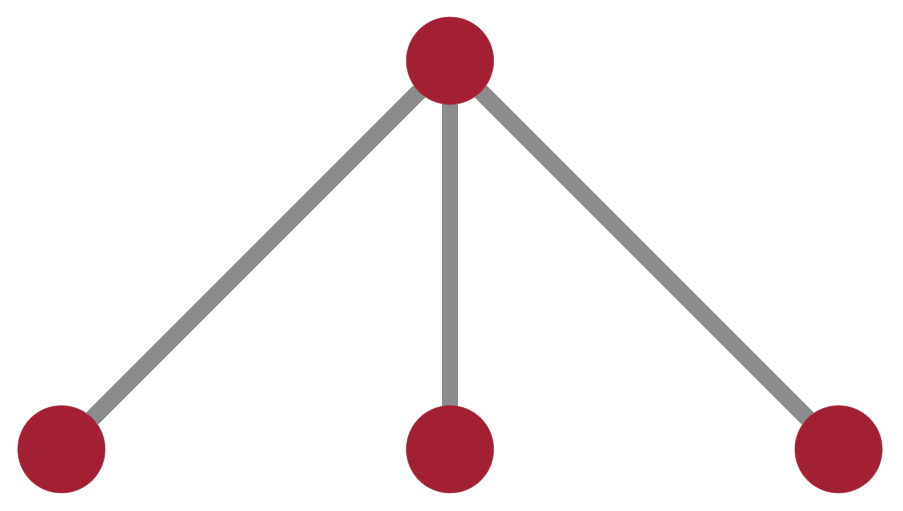}\end{gathered}$&=&
$\displaystyle\sum_{i_1=1}^M\sum_{i_2=1}^M\sum_{i_3=1}^M\sum_{i_4=1}^M z_{i_1}z_{i_2}z_{i_3}z_{i_4}  \theta_{i_1i_2}\theta_{i_1i_3}\theta_{i_1i_4}$\\
$d=3$ & $\begin{gathered}\includegraphics[scale=0.12]{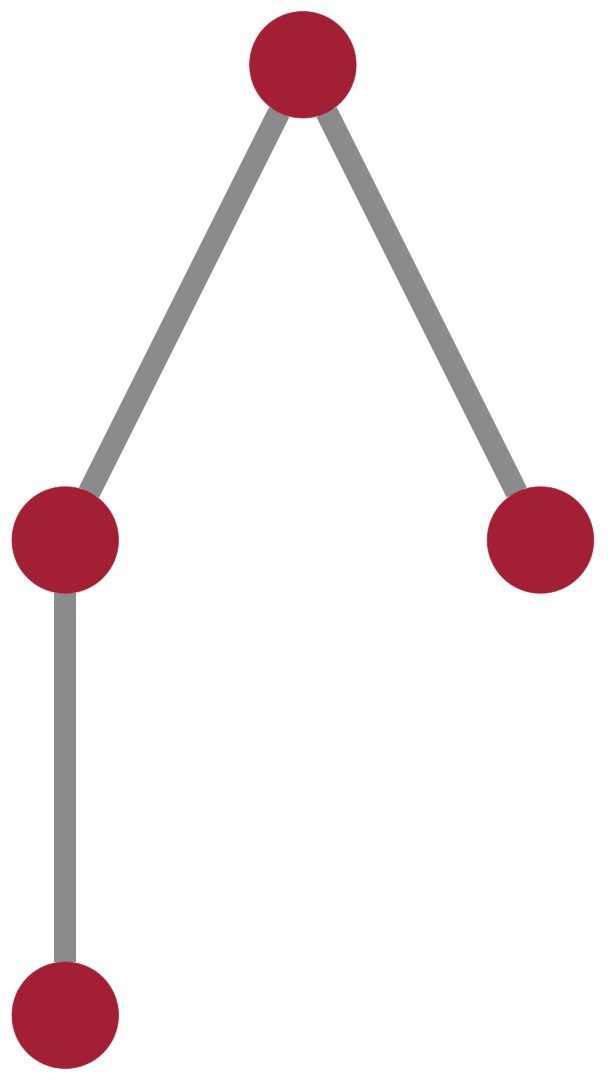}\end{gathered}$&=&
$\displaystyle\sum_{i_1=1}^M\sum_{i_2=1}^M\sum_{i_3=1}^M\sum_{i_4=1}^M z_{i_1}z_{i_2}z_{i_3}z_{i_4} \theta_{i_1i_2}\theta_{i_2i_3}\theta_{i_3i_4}$\\
&$\begin{gathered}\includegraphics[scale=0.12]{graphs/2_1_1.pdf}\end{gathered}
\begin{gathered}\includegraphics[scale=0.12]{graphs/2_2_1.pdf}\end{gathered}$&=&
$\displaystyle\sum_{i_1=1}^M\sum_{i_2=1}^M\sum_{i_3=1}^M\sum_{i_4=1}^M z_{i_1}z_{i_2}z_{i_3}z_{i_4} \theta_{i_1i_2} \theta_{i_3i_4}^2$\\
& $\begin{gathered} \includegraphics[scale=0.12]{graphs/2_1_1.pdf}\end{gathered}
\begin{gathered}\includegraphics[scale=0.12]{graphs/3_2_1.pdf}\end{gathered}$&=&
$\displaystyle\sum_{i_1=1}^M\sum_{i_2=1}^M\sum_{i_3=1}^M\sum_{i_4=1}^M\sum_{i_5=1}^M z_{i_1}z_{i_2}z_{i_3}z_{i_4}z_{i_5} \theta_{i_1i_2} \theta_{i_3i_4}\theta_{i_3i_5}$\\
&  $\begin{gathered}\includegraphics[scale=0.12]{graphs/2_1_1.pdf}\,
\includegraphics[scale=0.12]{graphs/2_1_1.pdf}\,
\includegraphics[scale=0.12]{graphs/2_1_1.pdf}\end{gathered}$&=&
$\displaystyle\sum_{i_1=1}^M\sum_{i_2=1}^M\sum_{i_3=1}^M\sum_{i_4=1}^M\sum_{i_5=1}^M\sum_{i_6=1}^M z_{i_1}z_{i_2}z_{i_3}z_{i_4}z_{i_5}z_{i_6} \theta_{i_1i_2}\theta_{i_3i_4}\theta_{i_5i_6}$\\
& & & \\
\hline
\hline
\end{tabular}
\caption{All distinct multiparticle correlators with degree $d\le 3$, shown as multigraphs and written explicitly.}
\label{tab:corrlist}
\end{table*}

\subsection{Kinematic Polynomials}
\label{sec:kinpoly}

As one application in high-energy physics, multiparticle correlators arise in the study of kinematic polynomials.
These are symmetric polynomials built from Mandelstam invariants:
\begin{equation}
s_{ij} = (p_i^\mu + p_j^\mu)^2 = 2 \, p_i^\mu p_{j\mu},
\end{equation}
where the last equality assumes massless particles ($p^2 = 0$).
Kinematic polynomials have been used to build operator bases~\cite{Hogervorst:2014rta,Henning:2017fpj} for quantum field theories and understand the structure of scattering amplitudes~\cite{Boels:2013jua}.

In our notation, kinematic polynomials are multiparticle correlators of the form:
\begin{align}
z_i &=1,\\
\theta_{ij} &= s_{ij},\label{eq:kinpolytheta}
\end{align}
and so the insights we develop here are directly applicable to this context.

It is worth noting that an analog of the graphical notation for multiparticle correlators was also developed in \Ref{Hogervorst:2014rta} and mentioned in \Ref{Henning:2017fpj}.
Beyond this, \Ref{Hogervorst:2014rta} remarks that there are linear relations among the kinematic polynomials which can complicate certain analyses, and that a classification of those relations would be required to pursue particular strategies.
We will explore these relations among multiparticle correlators at length, seeking to understand and classify these relations.

\subsection{Energy Flow Polynomials}
\label{sec:efps}

A family of multiparticle correlators relevant for collider physics is the set of EFPs~\cite{Komiske:2017aww}.
Explicitly, EFPs are defined as multiparticle correlators of the form:
\begin{align}
z_i &=E_i,\label{eq:efp1}\\
\theta_{ij} &= (2n_i^\mu n_{j\mu})^{\beta/2},\quad n_i^\mu \equiv p_i^\mu/E_i,\label{eq:efp2}
\end{align}
where $E_i$ is the energy of particle $i$, $n_i^\mu = (1,\hat n_i)^\mu$ for massless particles, and $\beta$ is an angular weighting factor.
While we use particle energies $E_i$ here to simplify the notation, in a hadron collider context these would be replaced with particle transverse momenta $p_{Ti}$.

The EFPs are IRC-safe observables, which are guaranteed to be finite and computable in perturbative quantum field theory~\cite{sterman1995handbook,Weinberg:1995mt,Kinoshita:1962ur,Lee:1964is,Frye:2018xjj}.
An observable is IRC safe if it is unchanged by the addition of a soft particle with $E\to0$ or by the collinear splitting of one particle into two with $p^\mu\to \{\lambda p^\mu, (1-\lambda)p^\mu\}$ for any $\lambda \in [0,1]$.
While \Ref{Komiske:2017aww} shows via direct computation that the EFPs are IRC safe, we will later develop a simple notation which makes this fact manifest in the $\beta=2$ case (see \Sec{subsec:revisit_efps}).

For any choice of $\beta > 0$, the EFPs form a linear basis of all IRC-safe observables~\cite{Komiske:2017aww}, meaning that any IRC-safe observable can be approximated by a finite linear combination of EFPs.
Many common collider observables are encompassed by the EFPs~\cite{Larkoski:2013eya,Moult:2016cvt} or can be cast as exact linear combinations of EFPs~\cite{Berger:2003iw,Almeida:2008yp,Ellis:2010rwa,Larkoski:2014uqa,Larkoski:2014pca}.
While it has been shown that the EFP basis can be useful for jet classification~\cite{Komiske:2017aww,Kasieczka:2019dbj}, the basis is overcomplete even for fixed $\beta$, in the sense that there are linear relations among the elements, which complicates the use of linear fitting methods.
Understanding these relations is a key goal of this paper, which we will accomplish in \Secs{subsec:vertex_linear}{subsec:edge_linear}.

\subsection{Computational Complexity}
\label{sec:complexity}

An $N$-particle correlator in \Eq{eq:multcorrG} is naively computable in $\mathcal O(M^N)$ by evaluating the $N$ nested sums over $M$ particles.
This computational complexity can typically be significantly improved upon, however, by exploiting the algebraic structure of the sums, as we have detailed in \Ref{Komiske:2017aww} and summarize here.
Efficiently computing multiparticle correlators is especially important for collider physics applications, where hundreds to thousands of final-state particles are produced in each collision.

The key insight is to iteratively perform the sums in a carefully chosen order.
For example, we can write the following correlator in a suggestive way:
\begin{align}
\label{eq:vecorr}
\begin{gathered}\includegraphics[scale=0.2,angle=90]{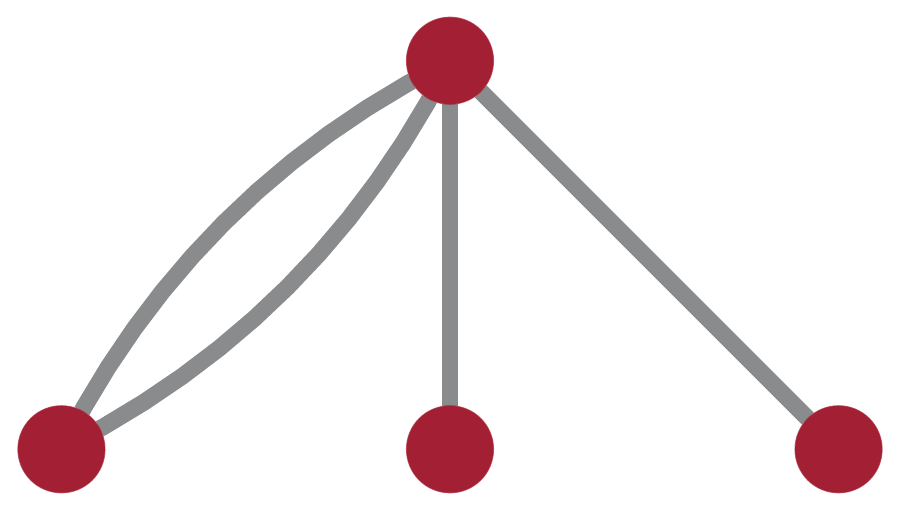}\end{gathered}
&=\sum_{i_1=1}^M\sum_{i_2=1}^M\sum_{i_3=1}^M\sum_{i_4=1}^M z_{i_1}z_{i_2}z_{i_3}z_{i_4}\theta_{i_1i_2}\theta_{i_1i_3}\theta_{i_1i_4}^2 \nonumber
\\& = \sum_{i_1=1}^M z_{i_1}\left( \sum_{i_2=1}^M z_{i_2} \theta_{i_1i_2}\right)^2\left( \sum_{i_4=1}^M  z_{i_4}\theta_{i_1i_4}^2\right),
\end{align}
which can be evaluated in $\mathcal O(M^2)$ by computing the parenthetical quantities in $\mathcal O(M)$ for all $M$ terms in the outer sum.
This strategy gives a significant improvement over the naive $\mathcal O(M^4)$ complexity of the correlator.

The general procedure to determine the order in which to compute the nested sums is known as the variable elimination algorithm~\cite{zhang1996exploiting}.
Using variable elimination yields a computational complexity of $\mathcal O(M^{\text{tw}(G)+1})$, where $\text{tw}(G)$ is the treewidth of $G$, neglecting multiple edges.
The treewidth measures how ``tree-like'' a graph is, with $\text{tw}(G)=1$ for trees and higher values for other graphs.
For all non-complete graphs, $\text{tw}(G)+1 < N$ and so variable elimination provides a significant improvement in computational speed in most cases.
Here, we see the graphical notation is not only useful for enumerating all possible correlators, but it also provides a natural strategy for efficiently evaluating the correlators.
In the next two sections, we exploit this graphical language for further computational gains.

\section{Cutting Open Vertices}
\label{sec:verts}

In this section, we begin to surgically disassemble the multigraphs by slicing them through their vertices.
This yields ``particle tensors'' whose indices index particles and allows us to view the correlators as contractions over these indices.
Using these particle tensors, we can understand linear relations that emerge when the number of particles is small as well as speed up certain computations beyond what is possible through variable elimination alone.
These results hold for any multiparticle correlator of the form in \Eq{eq:multcorrG}, regardless of the choice of pairwise distances $\theta_{ij}$.

\subsection{Particle Tensors}

One can view the sums over particle indices $i_1,\ldots,i_N$ in \Eq{eq:multcorrG} as contractions of tensorial objects.
We can explicitly find these tensors by slicing open the multigraph vertices.
For example, the following correlator can be written suggestively as:
\begin{align}
\asymwedgeA{0.16} &= \sum_{i_1=1}^M \sum_{i_2=1}^M\sum_{i_3=1}^M z_{i_1}z_{i_2}z_{i_3} \theta_{i_1i_2}^3 \theta_{i_2 i_3} \nonumber \\
&= \mathcal{T}_{i_1} \mathcal{T}^{(3)}_{i_1i_2} \mathcal{T}^{(1)}_{i_2 i_3} \mathcal{T}_{i_3},
\label{eq:cutvertex}
\end{align}
where the indices $i_1,i_2,i_3$ are contracted via Einstein summation notation, and we have defined the particle tensors as:
\begin{equation}
\mathcal{T}_{i} = \sqrt{z_i}, \qquad \mathcal{T}^{(\alpha_{12})}_{i_1 i_2} = \sqrt{z_{i_1}z_{i_2}} \, \theta^{\alpha_{12}}_{i_1 i_2}.
\end{equation}

Note that this decomposition is not unique, and we could have represented \Eq{eq:cutvertex} alternatively as:
\begin{equation}
\label{eq:cutvertex_alt}
\mathcal{T}_{i_1} \mathcal{T}_{i_2} \mathcal{T}_{i_3} \mathcal{T}^{(3,0,1)}_{i_1 i_2 i_3},
\end{equation}
where the 3-index particle tensor is:
\begin{equation}
\mathcal{T}^{(\alpha_{12},\alpha_{13},\alpha_{23})}_{i_1i_2i_3} = \sqrt{z_{i_1}z_{i_2} z_{i_3}} \, \theta^{\alpha_{12}}_{i_1 i_2} \theta^{\alpha_{13}}_{i_1 i_3} \theta^{\alpha_{23}}_{i_2 i_3}.
\end{equation}
This is an example of an $N$-index particle tensor:
\begin{equation}
\label{eq:generalNindex}
\mathcal{T}_{i_1\cdots i_N}^{(\{\alpha_{ab}\})} = \sqrt{\prod_{j=1}^N z_{i_j}} \prod_{k=1}^N\prod_{\ell=k+1}^N\theta_{i_k i_\ell}^{\alpha_{k\ell}},
\end{equation}
where $\{\alpha_{ab}\}$ are integers ordered by ascending $b$ then $a$.

The general decomposition, demonstrated by \Eq{eq:cutvertex}, has a simple graphical interpretation as graphs with cut vertices:
\begin{equation}
\begin{tikzpicture}[      
        every node/.style={anchor=south west,inner sep=0pt},
        x=1mm, y=1mm,
      ]   
     \node (fig2) at (3,3)
       {\includegraphics[scale=0.04]{graphs/1_0_1.pdf}};  
     \draw [fill=patchcolor,patchcolor] (3.25,5) rectangle (6.5,6.5);
\end{tikzpicture}_i = \sqrt{z_i}, \quad \quad \,_j\edgegraph{0.2}_k = \theta_{jk}.
\end{equation}
Here, we have sliced open a vertex into two free indices, which will be later contracted to obtain the sums.
For example, the alternative decomposition in \Eq{eq:cutvertex_alt} involves a graph with three cut vertices:
\begin{equation}
\begin{gathered}
\begin{tikzpicture}[      
        every node/.style={anchor=south west,inner sep=0pt},
        x=1mm, y=1mm,
      ]   
     \node (fig2) at (3,3)
       {\includegraphics[scale=0.16]{graphs/3_4_3.pdf}};  
     \draw [fill=patchcolor,patchcolor] (3,3) rectangle (5.5,4);
	\node[text width=0.05cm] at (1,1) {$i_1$} ;
     \draw [fill=patchcolor,patchcolor] (15.5,3) rectangle (18,4);
	\node[text width=0.05cm] at (18,1) {$i_3$} ;
     \draw [fill=patchcolor,patchcolor] (9.5,16.7) rectangle (11.5,17.5);
	\node[text width=0.05cm] at (9.5,17.5) {$i_2$} ;
\end{tikzpicture}\end{gathered}
 = \mathcal{T}_{i_1i_2i_3}^{(3,0,1)} = \sqrt{z_{i_1}z_{i_2}z_{i_3}} \, \theta_{i_1i_2}^3 \theta_{i_2i_3}.
\label{eq:examplehamburger}
\end{equation}
The generic $N$-index particle tensor in \Eq{eq:generalNindex} can be represented by a multigraph $G$:
\begin{equation}
\sqrt{z_{i_1}}\cdots \sqrt{z_{i_N}} \prod_{(k,\ell)\in G}\theta_{i_k i_\ell},
\end{equation}
which is similar to the multiparticle correlator in \Eq{eq:multcorrG}, albeit without the sums over particles and with square roots on the weights $z_i$.
We also note that the particle tensors, unlike the multiparticle correlators, depend on the particular vertex labeling of $G$.

For any multigraph, its vertices can be sliced in half in many different ways, each yielding a valid tensor contraction expression for that correlator.
All the different ways of cutting the vertices of the triangle graph yield:
\begin{align}\label{eq:slicetriangle}
\triangle{0.14} &= \sum_{i_1=1}^M\sum_{i_2=1}^M \sum_{i_3=1}^M z_{i_1}z_{i_2}z_{i_3} \theta_{i_1i_2}\theta_{i_2i_3}\theta_{i_1i_3}
\\&=\nonumber
 \begin{gathered}
\begin{tikzpicture}[      
        every node/.style={anchor=south west,inner sep=0pt},
        x=1mm, y=1mm,
      ]   
     \node (fig2) at (0,0)
       {\includegraphics[scale=0.14]{graphs/3_3_1.pdf}};  
     \node[text width=0.1cm,rotate=90] at (7.5,8.25) {\bf{|}};
	\node[text width=0.1cm, rotate=30] at (-0.2,-0.9) {\bf{|}} ;
	\node[text width=0.1cm, rotate=-30] at (9.6,1.2) {\bf{|}} ;
   ;
\end{tikzpicture}
\end{gathered}
= \mathcal{T}^{(1)}_{i_1 i_2} \mathcal{T}^{(1)}_{i_2 i_3} \mathcal{T}^{(1)}_{i_1 i_3}
\\&=\nonumber
 \begin{gathered}
\begin{tikzpicture}[      
        every node/.style={anchor=south west,inner sep=0pt},
        x=1mm, y=1mm,
      ]   
     \node (fig2) at (0,0)
       {\includegraphics[scale=0.14]{graphs/3_3_1.pdf}};  
     \node[text width=0.1cm,rotate=0] at (4.5,9.5) {\bf{|}};
	\node[text width=0.1cm, rotate=30] at (-0.2,-0.9) {\bf{|}} ;
	\node[text width=0.1cm, rotate=-30] at (9.6,1.2) {\bf{|}} ;
   ;
\end{tikzpicture}
\end{gathered}
= \mathcal{T}_{i_2}\mathcal T_{i_1i_2i_3}^{(1,0,1)} \mathcal T_{i_1i_3}^{(1)}
\\&=\nonumber
 \begin{gathered}
\begin{tikzpicture}[      
        every node/.style={anchor=south west,inner sep=0pt},
        x=1mm, y=1mm,
      ]   
     \node (fig2) at (0,0)
       {\includegraphics[scale=0.14]{graphs/3_3_1.pdf}};  
     \node[text width=0.1cm,rotate=0] at (4.5,9.5) {\bf{|}};
	\node[text width=0.1cm, rotate=120] at (3.1,-0.2) {\bf{|}} ;
	\node[text width=0.1cm, rotate=-30] at (9.6,1.2) {\bf{|}} ;
   ;
\end{tikzpicture}
\end{gathered}
\nonumber
= \mathcal{T}_{i_1}\mathcal{T}_{i_2}\mathcal T_{i_1i_2i_3i_3}^{(1,0,1,1,0,0)}
\\&=\nonumber
 \begin{gathered}
\begin{tikzpicture}[      
        every node/.style={anchor=south west,inner sep=0pt},
        x=1mm, y=1mm,
      ]   
     \node (fig2) at (0,0)
       {\includegraphics[scale=0.14]{graphs/3_3_1.pdf}};  
     \node[text width=0.1cm,rotate=0] at (4.5,9.5) {\bf{|}};
	\node[text width=0.1cm, rotate=120] at (3.1,-0.2) {\bf{|}} ;
	\node[text width=0.1cm, rotate=-120] at (12.2,3.4) {\bf{|}} ;
   ;
\end{tikzpicture}
\end{gathered}
\nonumber
= \mathcal{T}_{i_1} \mathcal{T}_{i_2} \mathcal{T}_{i_3} \mathcal{T}^{(1,1,1)}_{i_1i_2 i_3}.
\end{align}
The first contraction in \Eq{eq:slicetriangle} is the trace of the cube of a two-index tensor treated as a matrix, which we can graphically represent as:
\begin{equation}
\triangle{0.14}
 = \tr\left(
\begin{gathered}
\begin{tikzpicture}[      
        every node/.style={anchor=south west,inner sep=0pt},
        x=1mm, y=1mm,
      ]   
     \node (fig42) at (3,3)
       {\includegraphics[scale=0.16]{graphs/2_1_1}};  
     \draw [fill=patchcolor,patchcolor] (3,3) rectangle (5.5,3.8);
	\node[text width=0.05cm] at (1,1) {} ;
     \draw [fill=patchcolor,patchcolor] (3,20) rectangle (5.5,21);
	\node[text width=0.05cm] at (1,21) {} ;
\end{tikzpicture}\end{gathered}^3
 \right).\label{eq:trmatrix}
\end{equation}

\subsection{Computational Complexity}
\label{sec:partcc}

Casting the particle sums in multiparticle correlators as tensor contractions can be used to improve their computation in several ways.
First, improvements in the computational complexity of individual correlators can be achieved using algorithms for fast matrix operations.
While the triangle graph in \Eq{eq:slicetriangle} is $\mathcal O(M^3)$ to evaluate both naively and with variable elimination, casting it as a matrix product in \Eq{eq:trmatrix} allows us to go beyond this limit.
Using Strassen's algorithm~\cite{strassen1969gaussian} yields an $\mathcal O(M^{2.81})$ evaluation of the correlator, and methods as fast as $\mathcal O(M^{2.38})$ exist~\cite{DBLP:journals/jsc/CoppersmithW90,davie2013improved,DBLP:conf/issac/Gall14a}.
For instance, Strassen's algorithm uses the structure of the matrix product to reduce the number of operations required to multiply $2\times2$ block matrices to 7 rather than 8, which gives rise to a power of $\log_2 7 \simeq 2.81$ in its complexity when recursively applied.
We will showcase this speedup for computing collider observables in \Sec{sec:speedingcolliderobs}.

Beyond improving individual correlator evaluation, the constituent particle tensors can be used to improve the computation of collections of correlators.
Many correlators can be built out of the same particle tensors contracted in different ways.
This allows for a dynamic programming approach where various particle tensors are computed and then re-used many times to evaluate additional correlators. 
While determining the optimal set of subgraphs to compute for a target collection of correlators is beyond the scope of this work, we note that it is in an interesting avenue for further exploration.

\subsection{Finite Particle Linear Relations}
\label{subsec:vertex_linear}

When there are small numbers of particles, algebraic relations among the multiparticle correlators emerge due to the simplified nature of the configuration.
Relations of this type were also explored in \Ref{Boels:2013jua} for the purpose of expanding superstring amplitudes.
In the collider context, understanding these relations is also important for calculating EFPs in fixed-order perturbative quantum field theory with small numbers of particles.
In general, understanding these finite-particle relations can improve the computation of collections of correlators by avoiding the computation of redundant elements.
Further, these relations can involve multigraphs of different computational complexities, allowing for individual correlators to be evaluated more efficiently via these relations.

The tensor identity in \Eq{eq:antisymid} concisely encodes many well-known results, including several Riemann tensor identities as well as the Cayley-Hamilton theorem~\cite{Edgar:2001vv}.
The second fundamental theorem of invariant theory indicates that any identity among a set of tensors in $m$ dimensions, aside from ones governed by their existing symmetries, can be obtained as a consequence of \Eq{eq:antisymid}~\cite{sneddon1998identities}.

\begin{table}[t]
\begin{tabular}{r @{$\qquad$} l}
\hline
\hline
Antisym. & Identity for $M = 2$ \\
\hline
\hline
$\begin{gathered}
\begin{tikzpicture}[      
        every node/.style={anchor=south west,inner sep=0pt},
        x=1mm, y=1mm,
      ]   
     \node (fig2) at (0,0)
       {\includegraphics[scale=0.14]{graphs/3_2_1.pdf}};  
     \node[text width=0.1cm] at (6.025,10.15) {\bf{]}};
	\node[text width=0.1cm, rotate=60] at (2.4,-0.25) {\bf{]}} ;
	\node[text width=0.1cm, rotate=-60] at (10.1,0.7) {\bf{]}} ;
   ;
\end{tikzpicture}
\end{gathered}$
&
$0 = 2\begin{gathered}\includegraphics[scale=0.14]{graphs/3_2_1.pdf}\end{gathered}  - \begin{gathered}\includegraphics[scale=0.14]{graphs/2_2_1.pdf}  \end{gathered} \dotgraph{0.02} $
\\
$\begin{gathered}
\begin{tikzpicture}[      
        every node/.style={anchor=south west,inner sep=0pt},
        x=1mm, y=1mm,
      ]   
     \node (fig2) at (0,0)
       {\includegraphics[scale=0.14]{graphs/3_3_1.pdf}};  
     \node[text width=0.1cm] at (6.025,8.95) {\bf{]}};
	\node[text width=0.1cm, rotate=30] at (1.4,-0.6) {\bf{]}} ;
	\node[text width=0.1cm, rotate=-30] at (10.65,0) {\bf{]}} ;
   ;
\end{tikzpicture}
\end{gathered}$
&
$0 = \begin{gathered}\includegraphics[scale=0.14]{graphs/3_3_1.pdf}\end{gathered}$
\\
$\begin{gathered}
\begin{tikzpicture}[      
        every node/.style={anchor=south west,inner sep=0pt},
        x=1mm, y=1mm,
      ]   
     \node (fig2) at (0,0)
       {\includegraphics[scale=0.14]{graphs/2_3_1.pdf}};
	\node (fig3) at (5,6.5)
	 {\includegraphics[scale=0.02]{graphs/1_0_1.pdf}};  
     \node[text width=0.1cm] at (5.45,5.8) {\bf{]}};
	\node[text width=0.1cm, rotate=-10] at (1.62,-0.6) {\bf{[}} ;
	\node[text width=0.1cm, rotate=-12] at (1.7,13) {\bf{]}} ;
   ;
\end{tikzpicture}
\end{gathered}$
&
$0 =\begin{gathered}\includegraphics[scale=0.14]{graphs/2_3_1.pdf}\end{gathered}  \dotgraph{0.02} 
- 2 \begin{gathered}\includegraphics[scale=0.14]{graphs/3_3_2.pdf}\end{gathered}$
\\
$\begin{gathered}
\begin{tikzpicture}[      
        every node/.style={anchor=south west,inner sep=0pt},
        x=1mm, y=1mm,
      ]   
     \node (fig2) at (0,0)
       {\includegraphics[scale=0.14]{graphs/4_3_2.pdf}};
     \node[text width=0.1cm, rotate=120] at (9.6,8.1) {\bf{]}};
	\node[text width=0.1cm, rotate=-90] at (-0.6,1.7) {\bf{]}} ;
	\node[text width=0.1cm, rotate=-100] at (-0.45,8.75) {\bf{]}} ;
   ;
\end{tikzpicture}
\end{gathered}$
&
$0 =2 \,\begin{gathered}\includegraphics[scale=0.14]{graphs/4_3_2.pdf}\end{gathered}
-\begin{gathered}\includegraphics[scale=0.14]{graphs/2_2_1.pdf}\end{gathered}\begin{gathered}\includegraphics[scale=0.14]{graphs/2_1_1.pdf}\end{gathered}
-\begin{gathered}\includegraphics[scale=0.14]{graphs/3_3_1.pdf}\end{gathered} \dotgraph{0.02} $
\\
$\begin{gathered}
\begin{tikzpicture}[      
        every node/.style={anchor=south west,inner sep=0pt},
        x=1mm, y=1mm,
      ]   
     \node (fig2) at (0,0)
       {\includegraphics[scale=0.16]{graphs/4_3_1.pdf}};
     \node[text width=0.1cm, rotate=-20] at (6.3,6) {\bf{]}};
	\node[text width=0.1cm, rotate=40] at (1.75,-0.5) {\bf{]}} ;
	\node[text width=0.1cm, rotate=-90] at (5.8,1.65) {\bf{]}} ;
   ;
\end{tikzpicture}
\end{gathered}$
&
$0 = \begin{gathered}\includegraphics[scale=0.16]{graphs/4_3_1.pdf}\end{gathered}
+\begin{gathered}\includegraphics[scale=0.14]{graphs/4_3_2.pdf}\end{gathered}
-\begin{gathered}\includegraphics[scale=0.14]{graphs/3_3_2.pdf}\end{gathered} \dotgraph{0.02}$
\\
\hline
\hline
\end{tabular}
\caption{All linear relations with $M=2$ particles for $d\le 3$ multiparticle correlators, derived by antisymmetrizing the particle tensor indices.}
\label{tab:M2linrel}
\end{table}

For the case of particle tensors with $M$ particles, antisymmetrizing over any choice of $L>M$ indices must vanish.
We can translate this fact into the language of cutting graphs.
Graphically, we denote the antisymmetrization over an index by a bracket through a vertex whose direction indicates which of the two particle tensors to use, and the vertex itself represents contracting the indices.
The sum over all permutations of those indices, weighted by the sign of the permutation, must vanish by \Eq{eq:antisymid}.
After contracting the indices, each permutation gives rise to an associated multigraph, so this procedure gives rise to an alternating sum of multigraphs which must vanish.
Some choices of antisymmetrizations will vanish trivially by the symmetries of the tensors, but many do not trivially vanish and give rise to non-trivial algebraic relations.

To showcase this approach, in \Tab{tab:M2linrel} we enumerate all of the finite particle linear relations for the case of $M=2$ and $d\le 3$.
For example, the first row of \Tab{tab:M2linrel} corresponds to the relation:
\begin{equation}
3!\,\mathcal{T}_{[i_1}  \mathcal{T}^{i_1}_{i_2} \mathcal{T}^{i_2}_{i_3 ]} \mathcal{T}^{i_3} = 2 \, \mathcal{T}_{i_1}  \mathcal{T}^{i_1}_{i_2} \mathcal{T}^{i_2}_{i_3} \mathcal{T}^{i_3} -  \mathcal{T}^{i_1}_{i_2} \mathcal{T}^{i_2}_{i_1}   \mathcal{T}^{i_3} \mathcal{T}_{i_3},
\end{equation}
where we have raised and lowered indices to make the summation convention more clear.
This procedure can be extended to enumerate the relations for larger numbers of particles.

\section{Cutting Open Edges}
\label{sec:edges}

In this section, we show that the structure of multiparticle correlators can be dramatically simplified if the pairwise distances $\theta_{ij}$ can be written as an inner product.
In particular, this allows us to slice open the multigraph edges to obtain ``moment tensors'', resulting in further computational speedups and a greater understanding of various linear relations.

\subsection{Moments from an Inner Product}
\label{subsec:innerproduct}

The following results assume that $\theta_{ij}$ can be written as an inner product.
Repeating \Eq{eq:innerprod} for convenience:
\begin{equation}
\label{eq:inprod}
\theta_{ij} = \eta_{\mu\nu} u_i^\mu u_{j}^\nu,
\end{equation}
for some vectors $u_i^\mu,u_j^\nu$ and (symmetric) metric $\eta_{\mu\nu}$.
This assumption holds for the kinematic polynomials in \Eq{eq:kinpolytheta} and for the EFP angular measure in \Eq{eq:efp2} for $\beta=2$.
While we use the notation of Lorentz indices and the Minkowski metric here, the conclusions hold for any inner product among vectors using any choice of metric.

With this assumption, any multiparticle correlator can be written via tensor contractions, where in this case the indices are Lorentz indices rather than particle indices.
To that end, we introduce the following moment tensors:
\begin{equation}
\label{eq:pmom}
\mathcal M^{\mu_1\cdots \mu_v} = \sum_{i=1}^M z_i \, u_i^{\mu_1} \cdots u_i^{\mu_v},
\end{equation}
where $v$ is the rank of the tensor.
We apply this moment logic to kinematic polynomials in \Sec{subsec:revisit_kinepoly} and to $\beta = 2$ EFPs in \Sec{subsec:revisit_efps}.

A simple example demonstrates the connection between multiparticle correlators and these moment tensors.
Consider the following equivalent ways of writing the simplest 2-correlator:
\begin{align}
\begin{gathered}\includegraphics[scale=0.15, angle=90]{graphs/2_1_1}\end{gathered} & = \sum_{i_1=1}^M \sum_{i_2=1}^M z_{i_1} z_{i_2} (\eta_{\mu\nu}u_{i_1}^\mu  u_{i_2}^\nu)
\label{eq:mass2}
\\&=\left(\sum_{i_1=1}^Mz_{i_1} u_{i_1}^{\mu}\right)\left(\sum_{i_2=1}^Mz_{i_2} u_{i_2}^{\nu}\right)\eta_{\mu\nu}
\nonumber 
\\&= \mathcal M^\mu\mathcal M_\mu, \nonumber
\end{align}
where the first line uses the definition of a multiparticle correlator and \Eq{eq:inprod}, and the last line is explicitly a contraction of moment tensors from \Eq{eq:pmom} with Einstein summation notation.
This is the generalization of the familiar mass identity in \Eq{eq:massexample}, where the inner product allows us to separate the factors into tensors with only a single Lorentz index.

We now generalize this procedure to demonstrate that all multiparticle correlators can be obtained from these moment tensors when the inner product condition holds.
The general strategy is to regroup the summand of a general correlator into a product of factors, one for each sum index.
The resulting expression manifestly factors the dependence on the constituent particles into tensorial objects contracted according to the multigraph.
Substituting \Eq{eq:inprod} into \Eq{eq:multcorrG}, we have:
\begin{align}
\sum_{i_1=1}^M&\cdots\sum_{i_N=1}^M z_{i_1}\cdots z_{i_N} \prod_{(k,\ell) \in G}\eta_{\mu\nu}  u_{i_k}^\mu u_{i_\ell}^\nu\nonumber\\
& = \left( \prod_{j=1}^N\sum_{i_j=1}^M z_{i_j}  u_{i_j}^{\mu_1^j} u_{i_j}^{\mu_2^j}\cdots u_{i_j}^{\mu_{v_j}^j}\right)\prod_{(k,\ell)\in G} \eta_{\mu^{k}_{A_{k\ell}} \mu^{\ell}_{A_{\ell k}}}\nonumber\\
&=\left( \prod_{j=1}^N {\mathcal M}^{\mu_1^j \mu_2^j\cdots \mu_{v_j}^j}\right)  \prod_{(k,\ell)\in G} \eta_{\mu^{k}_{A_{k\ell}} \mu^{\ell}_{A_{\ell k}}}, 
\label{eq:corr2mom}
\end{align}
where $v_j$ is the valency of vertex $j$, $\mu^a_b$ is an index corresponding to the $b^{\text{th}}$ edge associated with vertex $a$, and $A$ is a matrix such that the $A_{k\ell}^{\text{th}}$ instance of $u_{i_k}^\mu$ is contracted with the $A_{\ell k}^{\text{th}}$ instance of $u_{i_\ell}^\mu$.

Demonstrating their continued usefulness, multigraphs provide a simpler recipe for translating between multiparticle correlators and moment tensors than the unavoidably opaque notation of \Eq{eq:corr2mom}.
Each vertex in the graph is associated with a moment having $v$ indices, where $v$ is the valency of the vertex.
The edges in the graph specify which vertices are connected, and hence they encode the contractions between indices of different moments.
These rules can be summarized as:
\begin{align}
\begin{gathered}
\begin{tikzpicture}[      
        every node/.style={anchor=south west,inner sep=0pt},
        x=1mm, y=1mm,
      ]   
     \node (fig2) at (3,3)
       {\includegraphics[scale=0.26]{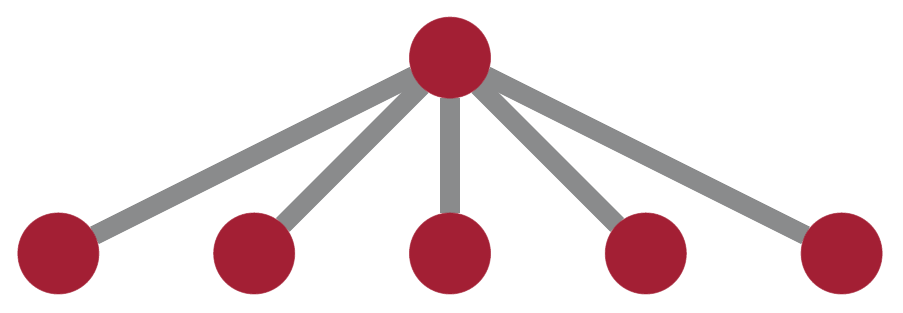}};  
     \draw [fill=patchcolor,patchcolor] (3,3) rectangle (26.5,5.75);
	\node[text width=0.05cm] at (3.75,3) {$^k$} ;
	\node[text width=0.05cm] at (13,3) {$\cdots$} ;
	\node[text width=0.05cm] at (24.25,3) {$^\ell$} ;
\end{tikzpicture}\end{gathered}  &= \mathcal M^{\mu_k \cdots \mu_\ell}\\
\label{eq:erule}\begin{gathered}_i\, \includegraphics[scale=0.21]{graphs/0_1_1}\,_j\end{gathered} &= \eta_{\mu_i \mu_j}.
\end{align}
Graphically, we have cut the edges of the multigraphs in half to obtain these tensors.

As an example, consider the following multigraph expressed both as a multiparticle correlator and in terms of contractions of moments:
\begin{align}
\label{eq:exefptoefm}
 \begin{gathered}
\begin{tikzpicture}[      
        every node/.style={anchor=south west,inner sep=0pt},
        x=1mm, y=1mm,
      ]   
     \node (fig2) at (0,0)
       {\includegraphics[scale=0.15]{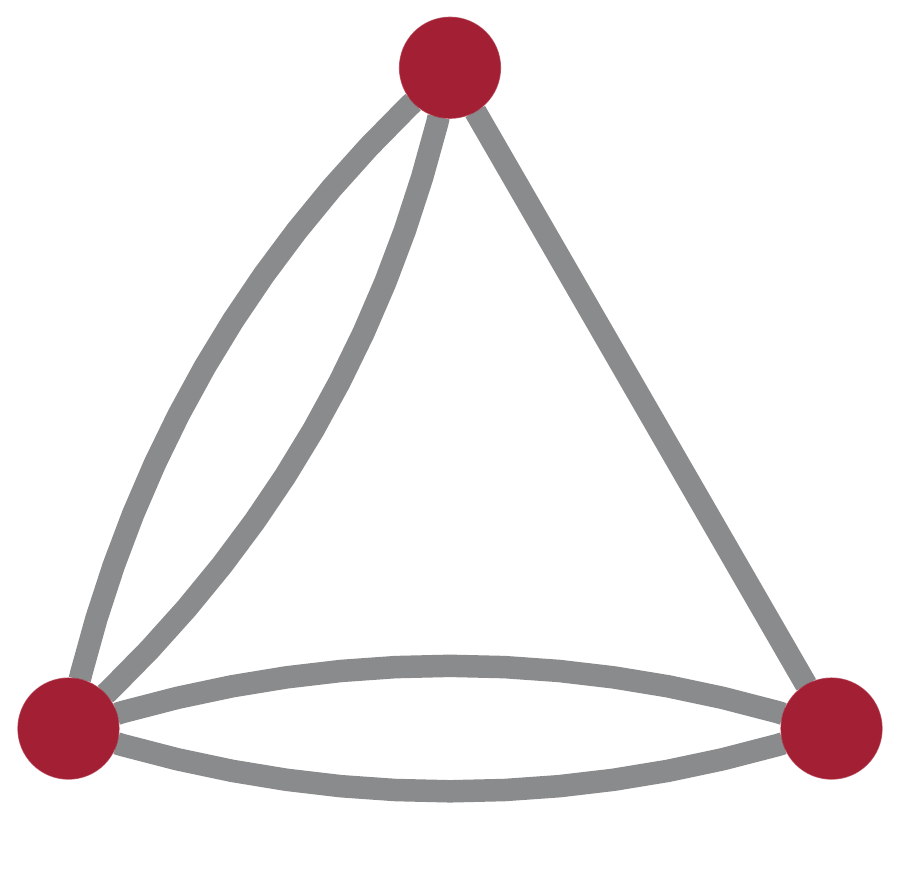}};  
     \node[text width=0.1cm,rotate=90] at (8,0) {\bf{|}};
	\node[text width=0.1cm, rotate=150] at (6.25,7) {\bf{|}} ;
	\node[text width=0.1cm, rotate=-150] at (11.25,9) {\bf{|}} ;
   ;
\end{tikzpicture}
\end{gathered}&= \sum_{i_1=1}^M\sum_{i_2=1}^M\sum_{i_3=1}^Mz_{i_1}z_{i_2}z_{i_3}\theta_{i_1i_2}\theta_{i_1i_3}^2\theta_{i_2i_3}^2\\
&=\mathcal M^{\mu\nu\rho\sigma}\mathcal M\indices{^\tau_{\mu\nu}} \mathcal M_{\rho\sigma\tau},
\end{align}
where pairs of indices are contracted with the metric.
Interestingly, these graphical rules are closely related to Penrose graphical notation for tensor contractions~\cite{penrose1971applications}.

\subsection{Computational Complexity}
\label{subsec:orderMcompute}

The moments in \Eq{eq:pmom} are symmetric tensor structures summed over $M$ particles, and hence are $\O(M)$ to compute.
Since the tensor structures are fully symmetric, they only have $(n+v-1)!/v!/(n-1)!$ independent components, where $n$ is the dimensionality of the inner product vectors $u^\mu_i$ and $v$ is the rank of the tensor.
The number of independent components is therefore polynomial in the rank $v$ as opposed to exponential.
In $3+1$ spacetime dimensions, $n=4$ and there are $(v+3)(v+2)(v+1)/6$ independent components.
Since this is independent of $M$ and $v^3$ is smaller than the number of particles for many events and multiparticle correlators of interest, computing the moment tensors is typically computationally efficient.

Since any multiparticle correlator with an inner product structure can be written as a contraction of moment tensors via \Eq{eq:corr2mom}, all such correlators become $\O(M)$ to compute.
The tensorial rearrangement into moments provides a way to circumvent the previous computational limits and achieve linear complexity.
This can be seen explicitly in \Eq{eq:mass2} where an $\O(M^2)$ tree graph is written as a contraction of two tensors that are each $\O(M)$ to compute.
Since the computation of any correlator must probe all particles at least once, the linear complexity in $M$ of computing the correlators via the moments is optimal.
Thus, by computing all of the particle moments up to a desired order $v$, all correlators can be obtained via tensor contractions which are independent of $M$ and depend only on the dimensionality of the space.
We provide an explicit demonstration of this computational improvement for collider observables in \Sec{sec:speedingcolliderobs}.

\subsection{Finite Dimension Linear Relations}
\label{subsec:edge_linear}

When the dimensionality of the inner product space is small, there are additional linear relations among the multiparticle correlators.
These relations were pointed out by \Ref{Hogervorst:2014rta}, and we now seek to understand them systematically.
Similar to the finite-particle relations in \Sec{subsec:vertex_linear}, the key is to figure out the appropriate way to apply \Eq{eq:antisymid} to antisymmetrize over more indices than dimensions.

In this case, we antisymmetrize over the inner product (Lorentz) indices instead of the particle indices.
We specialize to the cases of $n= 4$ and $n=3$ dimensions for the following discussion, since these are the dimensionalities relevant for \Sec{sec:energyflow} and \App{app:euclideanslicing}, respectively.
It is worth noting that $n$-dimensional identities continue to hold in fewer than $n$ dimensions.

The Cayley-Hamilton theorem is a special case of \Eq{eq:antisymid}, so we will first explore this as a method to understand the finite-dimension identities.
The Cayley-Hamilton theorem, as typically stated, says that a matrix $A$ satisfies its own characteristic polynomial.
Using the Newton identities, this translates to a matrix relation between powers of $A$, the trace of powers of $A$, and the determinant of $A$.
Written out explicitly for small matrices, we have for $2\times2$ matrices:
\begin{equation}\label{eq:CH22}
A^2 - (\tr A) A + (\det A) I=0,
\end{equation}
where $I$ is the identity matrix, and for $3\times 3$ matrices:
\begin{equation}\label{eq:CH33}
A^3 - (\tr A)A^2  + \frac{1}{2} \big( (\tr A)^2 - \tr A^2 \big) A- (\det A)I=0.
\end{equation}

For our first $n = 4$ identity, we apply the $4\times4$ Cayley-Hamilton theorem to the matrix
\begin{equation}
\sqrt{\eta}^{\mu_1\nu}\, \mathcal M_{\nu\rho} \sqrt{\eta}^{\rho\mu_2},
\end{equation}
where $\sqrt{\eta}$ is defined because $\eta$ is symmetric.

Multiplying the resulting Cayley-Hamilton relation by the matrix and taking a trace yields:
\begin{equation}
\label{eq:pentagon}
n \leq 4:  \quad
6\times
\begin{gathered}\includegraphics[scale=0.15]{graphs/5_5_1}\end{gathered}
-
5\times
\begin{gathered}\includegraphics[scale=0.15]{graphs/3_3_1}\end{gathered}
\begin{gathered}\includegraphics[scale=0.15]{graphs/2_2_1}\end{gathered} = 0,
\end{equation}
where we assume that $\mathcal M_{\mu}^\mu=0$, as in our cases of interest due to the masslessness of the particles, with the full expression being easily obtainable but unwieldy.
Note that \Eq{eq:pentagon} could also have been obtained directly from the antisymmetrization identity
\begin{equation}
\begin{gathered}
\begin{tikzpicture}[      
        every node/.style={anchor=south west,inner sep=0pt},
        x=1mm, y=1mm,
      ]   
     \node (fig2) at (0,0)
       {\includegraphics[scale=0.16]{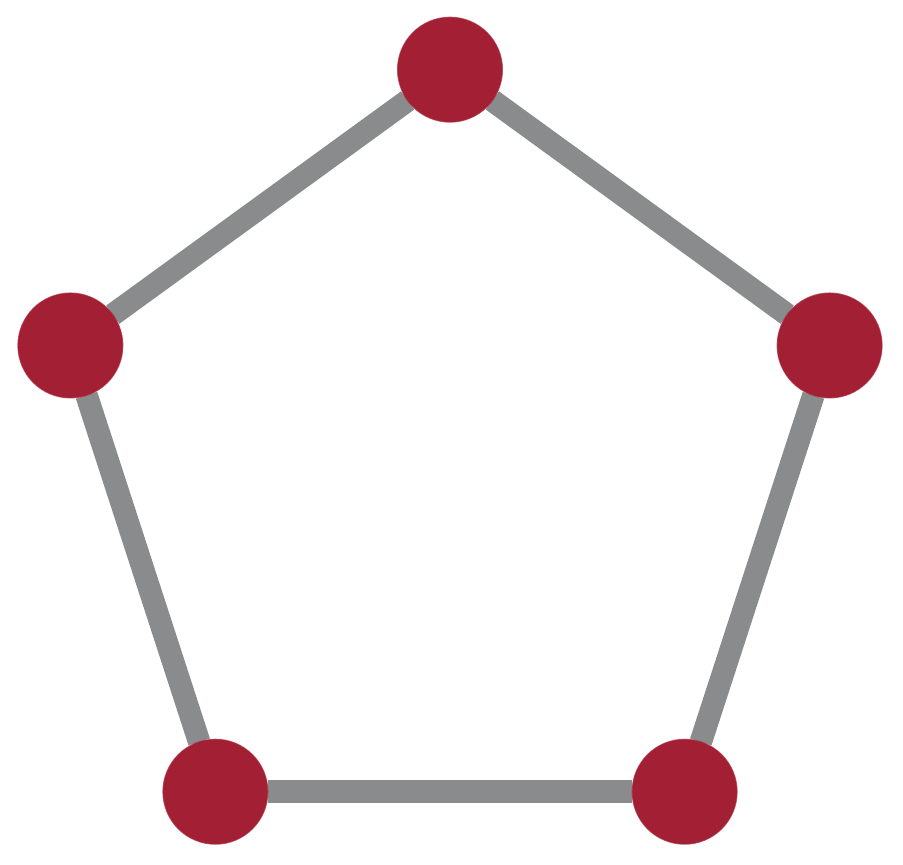}};  
     \node[text width=0.1cm] at (7 ,-0.5) {\bf{]}};
	\node[text width=0.1cm, rotate=72] at (14,3.75) {\bf{]}} ;
	\node[text width=0.1cm, rotate=144] at (12.25,11.9) {\bf{]}} ;
	\node[text width=0.1cm, rotate=216] at (3.75,12.6) {\bf{]}} ;
	\node[text width=0.1cm, rotate=288] at (0.5,4.75) {\bf{]}} ;
   ;
\end{tikzpicture}
\end{gathered}
=\mathcal M^{\mu_2}_{[\mu_1}\mathcal M^{\mu_3}_{\mu_2} \mathcal M^{\mu_4}_{\mu_3} \mathcal M^{\mu_5}_{\mu_4}\mathcal M^{\mu_1}_{\mu_5]}=0.
\end{equation}
Higher-dimensional versions of the Cayley-Hamilton theorem can be used to derive relations for larger 2-regular graphs (hexagon, heptagon, etc.).

In addition to identities involving 2-regular graphs, we can also multiply the expression of the Cayley-Hamilton theorem on both sizes by $\sqrt{\eta}^{\mu\nu}\mathcal M_\nu$ in order to produce relations among chain-like graphs, such as:
\begin{align}
\label{eq:bigchain}
n \leq 3:\quad   0&=
6\,\begin{gathered}\includegraphics[scale=0.15]{graphs/5_4_3}\end{gathered} 
-12\,\begin{gathered}\includegraphics[scale=0.15]{graphs/4_3_2}\end{gathered}
+6\,\begin{gathered}\includegraphics[scale=0.15]{graphs/2_1_1}\includegraphics[scale=0.15]{graphs/2_2_1}\end{gathered}
 +4\,\begin{gathered}\includegraphics[scale=0.15]{graphs/3_3_1}\end{gathered}
\nonumber\\ &\quad-2\,\begin{gathered}\includegraphics[scale=0.15]{graphs/3_3_1}\end{gathered}\begin{gathered}\includegraphics[scale=0.15]{graphs/2_1_1}\end{gathered}
-3\,\begin{gathered}\includegraphics[scale=0.15]{graphs/2_2_1}\end{gathered}\begin{gathered}\includegraphics[scale=0.15]{graphs/3_2_1}\end{gathered},
\end{align}
which holds in three or fewer dimensions.
In general, the Cayley-Hamilton theorem provides a matrix expression valid in some number of dimensions which can be combined with any graph fragment to give an identity.
While this approach only yields a subset of the finite-dimension identities, \Eq{eq:antisymid} captures all of them in full generality.

\section{Trimming the Leaves}
\label{sec:leaves}

When there are simple global constraints on the particles, their effects on multiparticle correlators can often be made clear graphically.
In this section, we consider cases where center-of-momentum relations allows us to ``trim the leaves'' of a multigraph, which gives us a way to count kinematic polynomials from \Ref{Boels:2013jua}.
More exotic graphical rules are relevant for multiparticle correlators in $e^+ e^-$ collisions, which we study in detail in \App{app:euclideanslicing}.

\subsection{Revisiting Kinematic Polynomials}
\label{subsec:revisit_kinepoly}

As anticipated in \Sec{subsec:innerproduct}, kinematic polynomials can be written as contractions of moment tensors.
With massless particles, the relevant moment tensors take the following form:
\begin{equation}
\mathcal P^{\mu_1 \cdots \mu_v} = 2^{v/2}\sum_{i=1}^M p_i^{\mu_1} \cdots p_i^{\mu_v},
\end{equation}
with the prefactor accounting for the factor of two in the Mandelstam invariants $s_{ij} = 2 p_i^\mu p_{j\,\mu}$.
Following the logic of \Sec{subsec:innerproduct}, all Lorentz-invariant kinematic polynomials can be built from complete Lorentz contractions of these moments.

\subsection{Center-of-Momentum Relations}

Kinematic polynomials can be further simplified when imposing energy-momentum conservation $\sum_i p_i^\mu = 0$.
Written in the generic multiparticle correlator language, this corresponds to the following constraint:
\begin{equation}
\mathcal M^\mu = \sum_{i=1}^Mz_i u_i^\mu =  0.
\end{equation}
Since $\mathcal M^\mu$ corresponds to a vertex connected to a single edge, we can write this constraint graphically as:
\begin{equation}
\label{eq:graphical_conservation}
\begin{tikzpicture}[      
        every node/.style={anchor=south west,inner sep=0pt},
        x=1mm, y=1mm,
      ]   
     \node (fig2) at (3,3)
       {\includegraphics[scale=0.16,angle=90]{graphs/2_1_1.pdf}};  
     \draw [fill=patchcolor,patchcolor] (13, 3) rectangle (21, 5);
\end{tikzpicture} \hspace{-7mm}= 0.
\end{equation}
Therefore, momentum conservation allows us to restrict our attention to ``leafless'' graphs, whose minimum vertex valency is greater than one.

Leafless graphs are also relevant for studying EFPs on a collection of final-state particles.
It is sometimes convenient to work in the center-of-momentum frame with the constraint
\begin{equation}
\mathcal M^\mu = (E,0,0,0)^\mu,
\end{equation}
where $E$ is a fixed energy in this frame, corresponding to the graphical rule:
\begin{equation}
\begin{tikzpicture}[      
        every node/.style={anchor=south west,inner sep=0pt},
        x=1mm, y=1mm,
      ]   
     \node (fig2) at (3,3)
       {\includegraphics[scale=0.16,angle=90]{graphs/2_1_1.pdf}};  
     \draw [fill=patchcolor,patchcolor] (13, 3) rectangle (21, 5);
\end{tikzpicture}
\hspace{-7mm}
{}^\mu
\quad
= 
\quad
     E
     \,
\delta_0^\mu.
\end{equation}
In the $e^+e^-$ case where $u^0_i = 1$, valency-one vertices simply contribute factors of $E$ and we have:
\begin{equation}
\mathcal M_{\mu_1}\mathcal M^{\mu_1\mu_2\cdots\mu_v} = E \mathcal M^{\mu_2\cdots\mu_v}.
\end{equation}
Hence only leafless multigraphs need to be considered in the center-of-momentum frame for EFPs in the case of $e^+e^-$ collisions.

\subsection{Counting Superstring Amplitudes}

There is an interesting dividend of our graphical understanding of the center-of-momentum relation in \Eq{eq:graphical_conservation}.
\Ref{Boels:2013jua} sought to count the number of independent, symmetric polynomials of kinetic variables of degree $d$, which is relevant for determining five-point superstring amplitudes.
They worked in the limit of many particles with many spacetime dimensions (i.e.~without finite spacetime dimension identities).
Using an interesting technique based on Molien series, they conjectured the counts for $d\le 8$.
In our language, counting these independent polynomials of degree $d$ in the center-of-momentum frame corresponds to simply counting leafless multigraphs with $d$ edges.

We can efficiently enumerate and count the number of leafless multigraphs using \textsc{Nauty}~\cite{McKay201494}, with our results summarized in \Tab{tab:sequence}.
Our sequence agrees with the results of \Ref{Boels:2013jua} where they overlap~\cite{oeisA226919}, but we are able to overcome previous computational limitations and double the number of known terms.
These values have been added as new sequences in the OEIS~\cite{oeisA307317,oeisA307316}.
Enumerating multigraphs not only allows us to efficiently count these polynomials, but it also provides explicit constructions of them, which may be useful in further exploring the space of kinematic polynomials.

\begin{table}[t]
\centering
\begin{tabular}{c @{$\quad$} r @{$\quad$} r}
\hline\hline
 & \multicolumn{2}{c}{\,Leafless Multigraphs} \\
 & Connected & All\\
Edges $d$ & \href{https://oeis.org/A307317}{A307317}~\cite{oeisA307317} & \href{https://oeis.org/A307316}{A307316}~\cite{oeisA307316} \\ \hline\hline 
1   &                 0 &                 0 \\
2   &                 1 &                 1 \\
3   &                 2 &                 2 \\
4   &                 4 &                 5 \\
5   &                 9 &                11 \\
6   &               26  &                34 \\
7   &               68  &                87 \\
8   &             217   &               279 \\
9   &             {\bf 718}   &         {\bf 897} \\
10  &        {\bf  2\,553}    &      {\bf 3\,129} \\
11  &        {\bf   9\,574}   &     {\bf 11\,458} \\
12  &       {\bf 38\,005}     &     {\bf 44\,576} \\
13  &     {\bf 157\,306}      &    {\bf 181\,071} \\
14  &    {\bf  679\,682}      &    {\bf 770\,237} \\
15  & {\bf 3\,047\,699}       & {\bf 3\,407\,332} \\
16  & {\bf 14\,150\,278} & {\bf 15\,641\,159} \\
\hline\hline
\end{tabular}
\caption{The number of multigraphs with $d$ edges that have no vertices of valency one (i.e.\ are leafless), tabulated for both connected and all graphs.
These have been added to the OEIS as new sequences~\cite{oeisA307317,oeisA307316}.
We conjecture that the sequence for all graphs is the same as \href{https://oeis.org/A226919}{A226919}~\cite{oeisA226919}, which was discovered in a string theory context~\cite{Boels:2013jua}.
The values for $d>8$ (bolded) are new results.}
\label{tab:sequence}
\end{table}

\section{Applications to Collider Physics}
\label{sec:energyflow}

In this section, we apply the general lessons that we have developed to obtain concrete computational and conceptual improvements for collider observables.%
\footnote{This section also explains two cryptic comments we made in previous papers: footnote 8 of \Ref{Komiske:2017aww} and footnote 4 of \Ref{Komiske:2018cqr}.}
We first introduce EFMs as novel tensorial structures to efficiently encode and compute $\beta = 2$ EFPs, and we show their relation to previous moment tensors in the literature.
We then exploit the results of \Sec{subsec:orderMcompute} to provide a maximally efficient $\mathcal O(M)$ implementation of key jet substructure observables.

\subsection{Introducing Energy Flow Moments}
\label{subsec:revisit_efps}

With $\beta=2$, the pairwise distance measure of the EFPs in \Eq{eq:efp2} has an inner product structure.
Following \Sec{subsec:innerproduct}, the $\beta=2$ EFPs can therefore be written as contractions of EFMs defined in \Eq{eq:efm}, repeated for convenience:
\begin{equation}
\label{eq:efm_again}
\mathcal I^{\mu_1 \cdots \mu_v} = 2^{v/2} \sum_{i=1}^M E_i \, n_i^{\mu_1} \cdots n_i^{\mu_v},
\end{equation}
where $n_i^\mu \equiv p_i^\mu/E_i$, and the prefactor accounts for the factor of two in $\theta_{ij}$.
The above definition is suitable for $e^+ e^-$ collisions, whereas in a hadron collider context, one would typically replace energy $E_i$ with transverse momentum $p_{Ti}$ in the definitions of both $\mathcal I^{\mu_1 \cdots \mu_v}$ and $n_i^\mu$.

The fact that the $\beta=2$ EFPs are IRC safe is immediately clear via inspection of \Eq{eq:efm_again}.
Due to the linear energy weighting, an EFM is manifestly invariant to the addition of a zero-energy particle or the collinear splitting of one particle into two.
To form the $\beta=2$ EFPs, we need only Lorentz contract the EFMs, which respects this IRC-safe structure.

As an aside, parity violation is interesting to explore at colliders~\cite{Nachtmann:1977ek,Dalitz:1988ab,Efremov:1992pe,Brandenburg:1995nv,Chatrchyan:2012dc,Aad:2015uau,Khachatryan:2016yte,Sirunyan:2018swq,Lester:2019bso}.
By invoking the $\epsilon$-symbol, we can construct parity-violating contractions of the EFMs that go beyond the parity-invariant EFPs.
Two simple parity-violating observables of potential interest are:
\begin{equation}
\label{eq:parity}
\epsilon_{\alpha\beta\gamma\delta}\I^{\alpha}\I^{\beta\rho}\I^{\gamma\sigma\tau}\I^\delta_{\rho\sigma\tau},\quad\epsilon_{\alpha\beta\gamma\delta}\I^{\alpha\rho}\I^{\beta\sigma}_\rho\I^{\gamma\tau}_\sigma\I^\delta_\tau,
\end{equation}
which are manifestly permutation symmetric, rotationally invariant, and IRC safe.
Further investigations in this direction are left to future work and we restrict ourselves here to contractions with the Minkowski metric.

\subsection{Relation to Exisiting Moments}
\label{subsec:existing_moments}

The EFMs are closely related to existing moment-based approaches for collider physics, dating back several decades.
Here, we show that EFMs directly encompass or suitably generalize these approaches.

A number of tensorial quantities similar in spirit to the EFMs have been previously defined and explored for both $e^+e^-$ and hadronic collisions.
In \Ref{Donoghue:1979vi}, the following objects built out of particle three-momenta were defined for $e^+e^-$ collisions:
\begin{equation}
\label{eq:thetadef}
\Theta^{j_1j_2\cdots j_v} = \sum_{i=1}^M E_i \,  \hat n_i^{j_1}  \hat n_i^{j_2} \cdots  \hat n_i^{j_v},
\end{equation}
where $ \hat n_i=\vec p_i/E_i$ is a unit three-vector and $j_k\in\{1,2,3\}$.
We refer to the tensors defined by \Eq{eq:thetadef} as ``generalized sphericity tensors'', since the second-rank tensor $\Theta_2=\Theta^{j_1j_2}$ is often called the sphericity tensor.
In an $e^+e^-$ context, the spatial components of the EFMs are exactly these tensors:
\begin{equation}
\mathcal I^{j_1j_2\cdots j_v}=2^{v/2}\,\Theta^{j_1j_2\cdots j_v},
\end{equation}
and, like the EFMs, the generalized sphericity tensors are IRC safe.

The second-rank sphericity tensor $\Theta_2=\Theta^{j_1j_2}$ has seen significant application in the form of defining observables from its eigenvalues.
The $C$ and $D$ parameter event shapes are defined in terms of the eigenvalues $\lambda_1$, $\lambda_2$, and $\lambda_3$ of the sphericity tensor $\Theta_2$ as:
\begin{equation}
\label{eq:cddef}
C = 3(\lambda_1\lambda_2 + \lambda_2\lambda_3 + \lambda_1\lambda_3), \quad D = 27 \lambda_1\lambda_2\lambda_3.
\end{equation}
These IRC-safe observables have been widely analyzed for $e^+e^-$ collisions both theoretically~\cite{Parisi:1978eg,Donoghue:1979vi,Ellis:1980wv,Catani:1998sf,Larkoski:2018cke} and experimentally~\cite{Ackerstaff:1997kk,Achard:2002kv,Heister:2003aj,Abdallah:2004xe,Bethke:2008hf}.

To highlight the close relationship with EFMs, we note that the $C$ and $D$ parameters are themselves both linear combinations of EFPs and contractions of EFMs:
\begin{align}\label{eq:Clor}
C  &= - \frac{3}{8} \begin{gathered}\includegraphics[scale=0.15]{graphs/2_2_1}\end{gathered} + \frac{3}{2}\,\,\begin{gathered}\includegraphics[scale=0.15]{graphs/2_1_1}\end{gathered}=-\frac{3}{8}\,\I^{\mu\nu}\I_{\mu\nu}+\frac32\,\I^\mu\I_\mu,\\
D &= -\frac{9}{8}\begin{gathered}\includegraphics[scale=0.15]{graphs/3_3_1}\end{gathered} + \frac{27}{4} \begin{gathered}\includegraphics[scale=0.15]{graphs/3_2_1}\end{gathered}-\frac{27}{8}\begin{gathered}\includegraphics[scale=0.15]{graphs/2_2_1}\end{gathered}\label{eq:Dlor}\\
&=-\frac98\,\I^{\mu\nu}\I_{\nu\rho}\I_{\mu}^\rho+\frac{27}{4}\,\I^\mu\I_{\mu\nu}\I^\nu-\frac{27}{8}\,\I^{\mu\nu}\I_{\mu\nu}\I.\nonumber
\end{align}
We derive these expressions in \App{app:euclideanslicing}.
In this graphical notation, we leave factors of total energy implicit:
\begin{equation}
\dotgraph{0.02} = \I = \sum_i E_i,
\end{equation} 
which must be included whenever graphs with different numbers of vertices are added together.

The third rank tensor $\Theta^{j_1j_2j_3}$ has seen additional study~\cite{Dagan:2011qy}, but the general formulation of \Eq{eq:thetadef} has seen very limited application at the LHC.
One reason for this may be that is is not clear how to use the tensors in \Eq{eq:thetadef} in a hadronic context: the factor of energy can be promoted to transverse momentum but the explicit appearance of the three-momenta prevents covariance under boosts and rotations about the collision axis.
The EFMs are precisely the generalization of the generalized sphericity tensors that applies in both the $e^+e^-$ and hadronic contexts.
Including the energy component of the particle four-momenta avails a hadronic interpretation simply by replacing all instances of energy with transverse momentum.
Relations among these $\Theta^{j_1j_2\cdots j_v}$ tensions are studied further in \App{app:euclideanslicing}.

Other constructions of moment tensors have been proposed solely in the hadronic context.
In \Ref{GurAri:2011vx}, a set of transverse-momentum-weighted moments in the rapidity-azimuth ($y,\phi$) plane were defined for narrow jets, and a number of substructure observables were identified in terms of their products and contractions.
Letting $x_i = (y_i,\phi_i)$ indicate the rapidity-azimuth position of particle $i$, and letting $k_j\in\{1,2\}$ be rapidity-azimuth indices, these hadronic moment tensors are:
\begin{equation}
\label{eq:hadrtensordef}
I^{k_1k_2\cdots k_v} = \sum_{i=1}^M p_{T,i} x_i\,^{k_1} x_i\,^{k_2} \cdots x_i\,^{k_v}.
\end{equation}
The two-index moment $I^{k_1k_2}$ has been used to define the planar flow observable~\cite{Thaler:2008ju,Almeida:2008tp}, which quantifies the linear versus planar nature of the radiation pattern.

When boosting a jet to be central ($y = 0$) and taking the narrow limit $y_i,\phi_i \ll 1$, $x_i$ become precisely two spatial indices of the EFMs with the hadronic prescription of $z_i = p_{T,i}$.
The tensors of \Eq{eq:hadrtensordef} were defined in \Ref{GurAri:2011vx} relative to a jet axis, which we neglect here (by boosting to $y=0$) to make the relation to EFMs immediate. 
Choosing a $p_T$-weighted centroid axis, a related analysis carries through with a jet axis included.
In this way, the EFMs generalize $I^{k_1\cdots k_n}$ beyond the narrow jet limit and do not require referencing a jet axis.

Thus, we see that the EFMs can be related to moments developed for both $e^+e^-$ and hadronic collisions.
Unlike these previous moments, which were developed for a specific type of collision, EFMs can operate in both realms by simply exchanging energies for transverse momenta.
In this way, EFMs are a general framework for moment-based approaches to collider physics analyses.

\subsection{Speeding Up Multi-Prong Taggers}
\label{sec:speedingcolliderobs}

Using the results of \Sec{subsec:orderMcompute}, EFPs with $\beta=2$ are manifestly $\mathcal O(M)$ to compute through the EFMs.
This provides a significant computational speedup over approaches based on variable elimination; see \Sec{sec:complexity}.

This computational speedup also applies to substructure observables derived from $\beta=2$ EFPs.
A number of different observables have been proposed to tag jets with multi-prong hadronic substructure at the LHC, including 2-pronged decays of boosted $W$ bosons and 3-pronged decays of boosted top quarks, along with other more exotic scenarios~\cite{Larkoski:2017jix,Marzani:2019hun}.
These jet substructure observables have since found significant experimental application at the LHC for multi-prong tagging and new physics searches~\cite{Asquith:2018igt}.
While most current LHC analyses focus on $\beta=1$ observables, comparable (or better) performance can be obtaining using $\beta = 2$ for standard tagging applications~\cite{Larkoski:2013eya,Larkoski:2014gra}.

As concrete examples, \Ref{Larkoski:2013eya} established a family of $N$-prong tagging observables $C_N^{(\beta)}$, which can be written as ratios of EFPs with complete graphs~\cite{Komiske:2017aww}.
For 2- and 3-prong tagging, these dimensionless ratio observables are:
\begin{equation}
C_2^{(\beta)} = \frac{\triangle{0.14}}{\linegraph{0.12}\quad \linegraph{0.12}}, \quad\quad C_3^{(\beta)} = \frac{\kite{0.14}\linegraph{0.12}}{\triangle{0.14}\triangle{0.14}},
\end{equation}
where factors of $\I = \dotgraph{0.02}$ can be restored by ensuring equal numbers of vertices in the numerator and denominator.
Note that these observables are traditionally defined using the rapidity-azimuth distance rather than \Eq{eq:efp2}, though these are equivalent for narrow jets and our $\mathcal O(M)$ moment logic can be applied in both cases.

These $C_N$ observables are in general $\mathcal O(M^{N+1})$ to compute, which becomes quickly computationally intractable with increasing $N$.
Using power counting arguments, \Ref{Larkoski:2014gra} established a different combination of energy correlators as an improved 2-pronged tagger called $D_2^{(\beta)}$, given by the expression:
\begin{equation}
D_2^{(\beta)} = \frac{\triangle{0.14}}{\linegraph{0.12}\quad \linegraph{0.12}\quad \linegraph{0.12}},
\end{equation}
which naively requires $\mathcal O(M^3)$ to compute.
These observables can be measured before or after grooming~\cite{Krohn:2009th,Ellis:2009me,Ellis:2009su,Dasgupta:2013ihk,Larkoski:2014wba} is applied, as they operate on generic sets of particles.
Note that the jet substructure observable $N_2$ cannot be expressed as an exact combination of EFPs, since the angular structure of the generalized correlators in \Ref{Moult:2016cvt} does not take the form of \Eq{eq:fdecomp}.

\begin{figure}[t]
\centering
\includegraphics[width=\columnwidth]{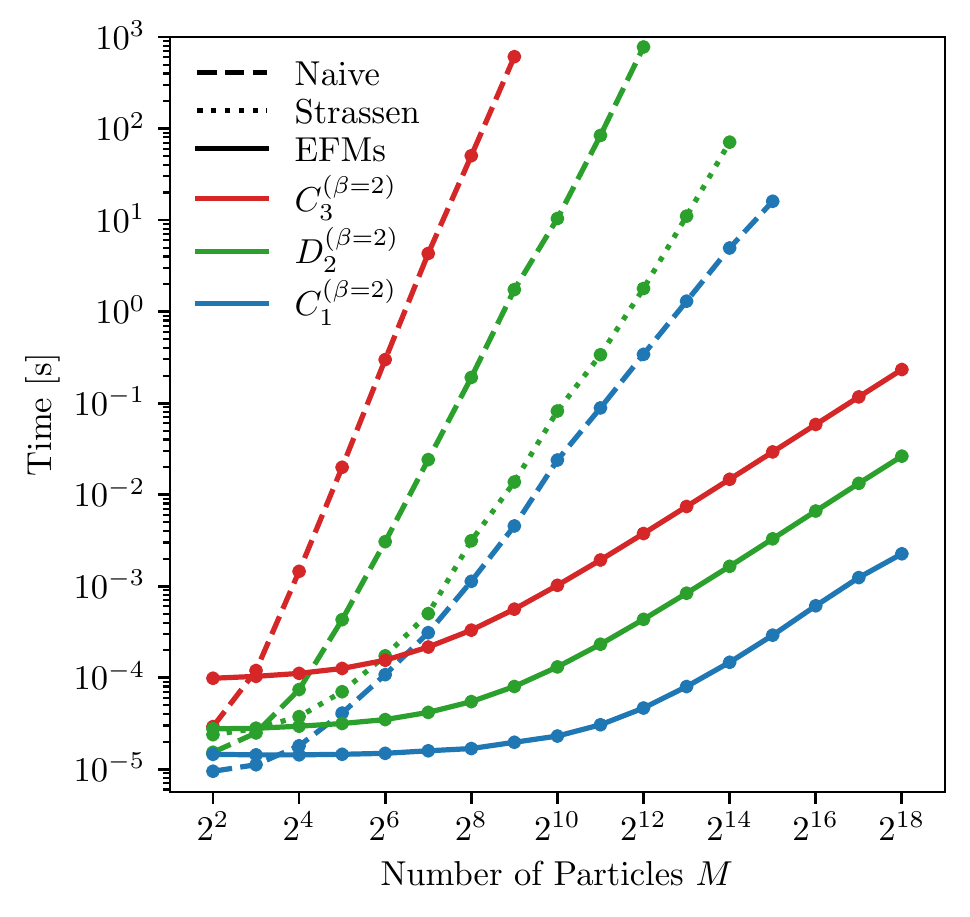}
\caption{
\label{fig:compspeed}
Compute time (in seconds) for three jet substructure observables: $C_3^{(\beta=2)}$ (naively $\O(M^4)$), $D_2^{(\beta=2)}$ (naively $\O(M^3)$), and $C_1^{(\beta=2)}$ (naively $\O(M^2)$) on an Intel Xeon 2.0 GHz processor.
The observables are computed for different numbers of particles by (dashed) evaluating the nested sums and (solid) using the moment-based approach developed here.
The fast matrix multiplication approach using Strassen's algorithm (dotted) is also used to compute $D_2$.
The moment strategy shows the expected linear scaling across all observables.
}
\end{figure}

In the case of $\beta=2$, these tagging observables can all be computed in $\mathcal O(M)$ using our moment-based approach.
Further, $C_2$ and $D_2$ benefit (for any $\beta$) from the fast matrix multiplication speedups to $\mathcal O(M^{2.81})$ discussed in \Sec{sec:partcc}.
In \Fig{fig:compspeed}, we compare the overall computational time of $C_3^{(\beta=2)}$ and $D_2^{(\beta=2)}$ using the naive approach and the moment-based approach.
We also show the fast matrix multiplication approach for computing $D_2$ using \textsc{BLAS}~\cite{blackford2002updated} via \textsc{NumPy}~\cite{numpycite,numpyweb} on eight CPU cores.
(The use of multiple cores merely shifts the curve vertically and does not alter the asymptotic scaling.)
For comparison, we include the computational time of a single 2-correlator (relevant for 1-prong quark/gluon discrimination~\cite{Larkoski:2014gra}):
\begin{equation}
C^{(\beta)}_1 =\linegraph{0.12} \,,
\end{equation}
which has a naive scaling of $\O(M^2)$.

We can see the significant benefit of linear computational complexity, resulting in very large practical speedups for realistic numbers of particles at the LHC (i.e.~100--1000).
Furthermore, this opens the door to the use of higher-$N$ correlators for collider physics, which until now have appeared computationally prohibitive.
The $\mathcal O(M)$ implementations of all these observables are available in the \href{https://energyflow.network}{\textsc{EnergyFlow}} Python package~\cite{EnergyFlow}.

\section{Conclusions}
\label{sec:conc}

A greater understanding of multiparticle correlators has profound implications for fields ranging from physics to applied mathematics.
In this paper, we have investigated multiparticle correlators, attempting to understand and categorize linear relations that appear among them as well as developing faster computational techniques to improve their practical application.

When the summand of a multiparticle correlator satisfies the monomial form in \Eq{eq:fdecomp}, the correlator has a multigraph structure that enables us to effectively represent and manipulate it.
This was used already in \Ref{Komiske:2017aww} to apply the variable elimination algorithm to speed up the computation of many multiparticle correlators for collider physics beyond the naive expectation.
Here, we have gone significantly further by developing techniques to ``cut'' the multigraph vertices and edges.

By slicing open the vertices of correlator graphs, we arrived at particle tensors and demonstrated that each correlator can be written as a contraction of these tensors.
The fundamental tensor antisymmetrization identity in \Eq{eq:antisymid} resulted in an infinite class of linear relations that hold when the number of particles $M$ is small compared to the number of vertices $N$.
The existence of faster-than-expected tensor operations, such as fast matrix multiplication, allowed us to reduce the computational complexity of correlators in general.
These computational and conceptual strategies extend beyond correlators with monomial summands, applying also to non-integer exponents or indeed products of any pairwise interparticle quantities.

We also sliced open the edges of the multigraphs when the pairwise distances had the inner product structure of \Eq{eq:innerprod}.
This assumption holds for kinematic polynomials as well as for EFPs with $\beta=2$ in both the $e^+e^-$ and hadron collider contexts.
Cutting the edges resulted in moment tensors that carry the indices of the inner product space.
Applying the tensor antisymmetrization identity led to another infinite class of linear relations arising when the dimensionality of the inner product space (e.g.~four spacetime dimensions) is small compared to the number of edges $d$.
We also made contact with efforts to analytically enumerate structures appearing in superstring amplitudes, deriving a method that allowed us to significantly extend previous results.

In a collider context, the moment tensors yield the EFMs.
The EFMs are novel tensorial structures that provide a natural and efficient way to compute collider correlators, unifying and encompassing existing moment-based approaches in the collider physics literature.
We demonstrated a significant speedup of widely used jet substructure observables using the graphical techniques that we developed.
We expect that the EFMs will be a useful development for experimental applications and theoretical calculations of multiparticle correlators in collider physics.

\begin{acknowledgments}
We are grateful to Brian Henning, Andrew Larkoski, Thomas Melia, and Ian Moult for helpful conversations.
We thank the Harvard Center for the Fundamental Laws of Nature for hospitality while this work was completed.
This work was supported by the Office of Nuclear Physics of the U.S. Department of Energy (DOE) under grant DE-SC-0011090 and by the DOE Office of High Energy Physics under grants DE-SC0012567 and DE-SC0019128.
JT is supported by the Simons Foundation through a Simons Fellowship in Theoretical Physics.
Cloud computing resources were provided through a Google Cloud allotment from the MIT Quest for Intelligence.
\end{acknowledgments}

\appendix

\section{Euclidean Slicing Linear Relations}
\label{app:euclideanslicing}

In this appendix, we derive linear relations for the generalized sphericity tensors in \Eq{eq:thetadef}, and show how they can be used to constrain $\beta = 2$ EFPs in $e^+e^-$ collisions.
The generalized sphericity tensors appear in the spatial parts of EFMs.
For example, the rank-2 tensors are related by:
\begin{equation}
\label{eq:rank2efmtheta}
\mathcal I^{00}=2,\quad\,\,\mathcal I^{0i}=\mathcal I^{i0}=\sqrt2\,\mathcal I^i=2\Theta^i,\quad\,\,\mathcal I^{ij}=2\Theta^{ij}.
\end{equation}
For massless particles in $e^+e^-$ collisions, the EFP measure in \Eq{eq:efp2} has the property that all $ n_i^\mu$ take the special form $ n_i^\mu=(1, \hat n)^\mu$.
The appearance of 1 in the zeroth component is the reason why relations such as $\mathcal I^{i0}=\sqrt2\,\mathcal I^i$ exist.
In general, we have ``subslicing'' relations of the form:
\begin{equation}
\mathcal I^{0\mu_1\cdots\mu_v}=\sqrt2\,\mathcal I^{\mu_1\cdots\mu_v}.
\end{equation}

Since the $\Theta$ tensors live in three dimensions, they satisfy finite-dimensional tensor identities involving antisymmetrization over four or more indices, i.e.\ one fewer than the four-dimensional EFMs.
Since any EFM for $e^+e^-$ collisions can be written in terms of the $\Theta$ tensors, these redundancies carry over and manifest as new EFM relations.
We showcase several specific examples of these identities in this appendix.

To begin, we introduce a graphical notation to keep track of the different contractions of the sphericity tensors, in analogy with the graphs for EFMs.
Explicitly, the graphical rules are:
\begin{align}
\begin{gathered}
\begin{tikzpicture}[      
        every node/.style={anchor=south west,inner sep=0pt},
        x=1mm, y=1mm,
      ]   
     \node (fig2) at (3,3)
       {\includegraphics[scale=0.26]{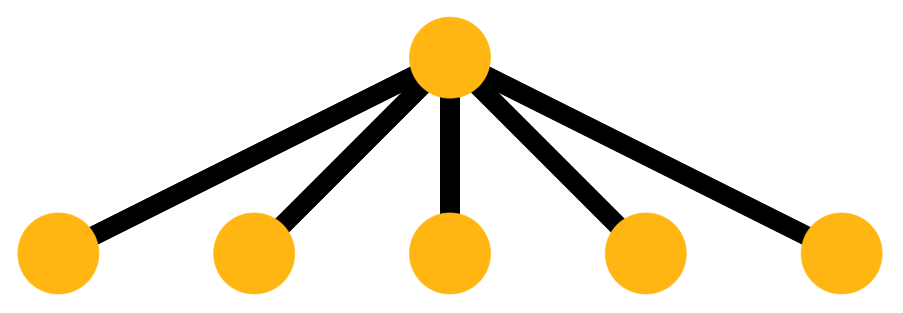}};  
     \draw [fill=patchcolor,patchcolor] (3,3) rectangle (26.5,5.75);
	\node[text width=0.05cm] at (3.75,3) {$^k$} ;
	\node[text width=0.05cm] at (13,3) {$\cdots$} ;
	\node[text width=0.05cm] at (24.25,3) {$^\ell$} ;
\end{tikzpicture}\end{gathered}  &= \Theta^{a_k \cdots a_\ell},\\
\label{eq:erule2}\begin{gathered}_i\,\includegraphics[scale=0.21]{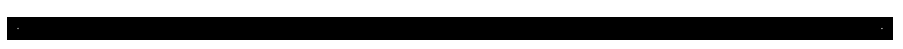}\,_j\end{gathered} &= 2\,\delta_{a_ia_j}.
\end{align}
For instance, the square graph has the following meaning in terms of the sphericity tensor:
\begin{equation}
\label{eq:eucgraphs}
\begin{gathered}\includegraphics[scale=0.14]{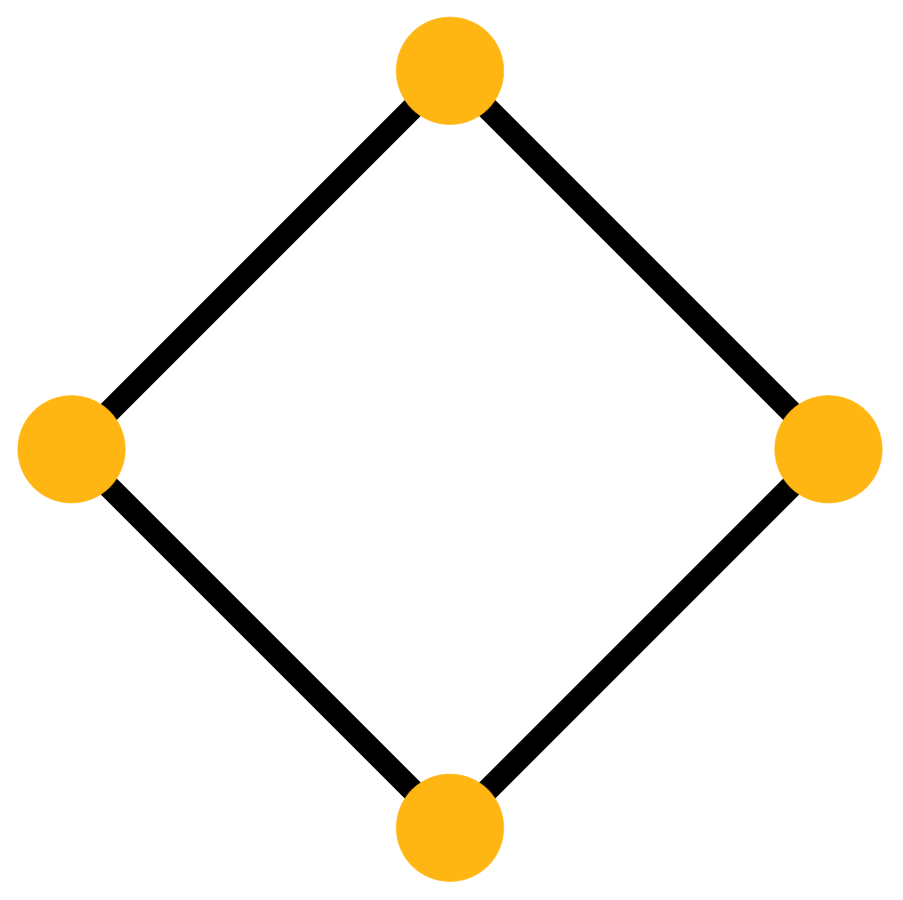}\end{gathered}
=2^4\, \Theta^{ab}\Theta_{bc}\Theta^{cd}\Theta_{da} =2^4\, \text{Tr}[\Theta_2^4].
\end{equation}
Note that $\text{Tr}[\Theta_2]=1$, and the factors of 2 come from the spatial part of our $2\delta_{ab}$ convention.
In general, the traces of powers of $\Theta_2$ give rise to 2-regular graphs.

\setlength{\tabcolsep}{8pt}
\begin{table*}[t]
\begin{tabular}{| >{\centering\arraybackslash} m{2em} || >{\centering\arraybackslash} m{2.5em} | >{\centering\arraybackslash} m{2.5em} | *{3}{>{\centering\arraybackslash} m{1.3em}} | *{8}{>{\centering\arraybackslash} m{1.4em}} | >{\centering\arraybackslash} m{1.5em}}\hhline{|~||*{14}{-}}
\multicolumn{1}{c|}{\unskip\textcolor{white!5}{\makebox[15pt]{\smash{\rule[8pt]{45pt}{5pt}}}} }  &\vspace{1mm} $d=0$ & $d=1$ &  \multicolumn{3}{c|}{$d=2$}  &  \multicolumn{8}{c|}{$d=3$}   &  \\
\multicolumn{1}{c|}{}  &  \includegraphics[scale=0.02]{graphs/1_0_1.pdf} & \includegraphics[scale=0.07]{graphs/2_1_1.pdf} & \includegraphics[scale=0.07]{graphs/2_2_1.pdf} & \includegraphics[scale=0.07]{graphs/3_2_1.pdf} &  \includegraphics[scale=0.07]{graphs/2_1_1.pdf} \includegraphics[scale=0.07]{graphs/2_1_1.pdf}  & \includegraphics[scale=0.07]{graphs/2_3_1.pdf} & \includegraphics[scale=0.07]{graphs/3_3_1.pdf} & \includegraphics[scale=0.07]{graphs/3_3_2.pdf}  & \includegraphics[scale=0.09]{graphs/4_3_1.pdf} & \includegraphics[scale=0.07]{graphs/4_3_2.pdf} & \includegraphics[scale=0.07]{graphs/2_1_1.pdf}\includegraphics[scale=0.07]{graphs/2_2_1.pdf} & \includegraphics[scale=0.07]{graphs/2_1_1.pdf}\includegraphics[scale=0.07]{graphs/3_2_1.pdf}&  \includegraphics[scale=0.07]{graphs/2_1_1.pdf} \includegraphics[scale=0.07]{graphs/2_1_1.pdf} \includegraphics[scale=0.07]{graphs/2_1_1.pdf}& $\cdots$
\\ \hhline{:-::*{14}{=}}\unskip\textcolor{white!5}{\makebox[-3pt]{\smash{\rule[12pt]{45pt}{4pt}}}} 
\includegraphics[scale=0.02]{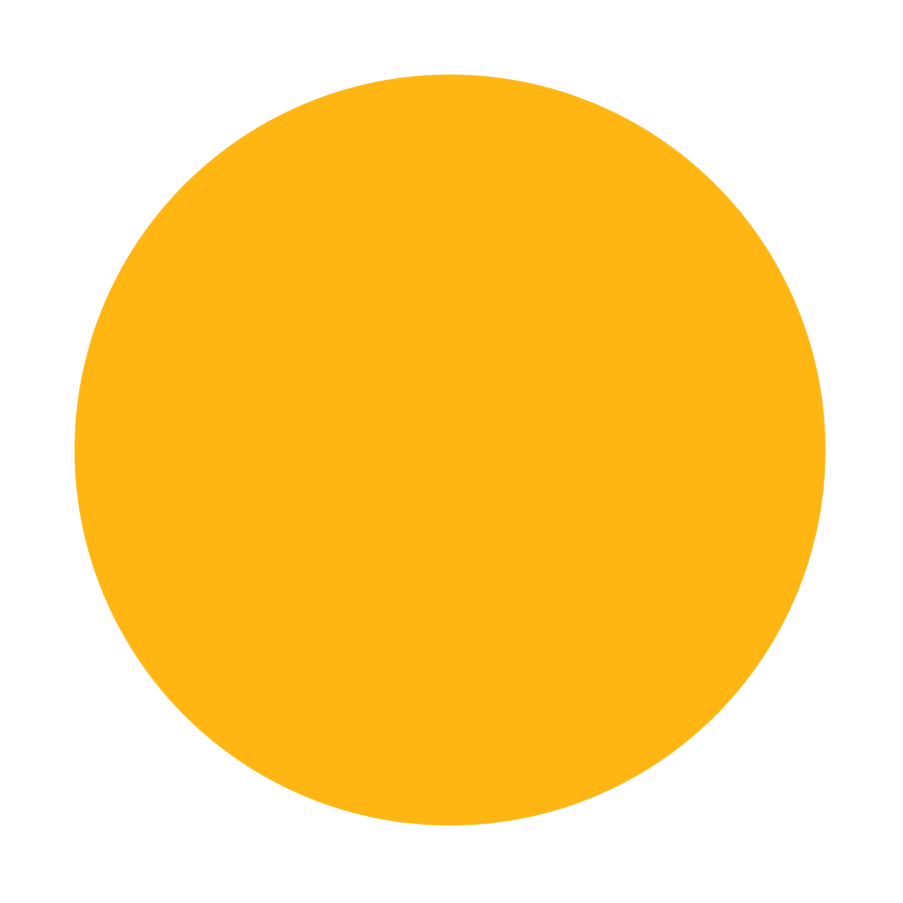} & 1 &  &  &  &   &  &  &   &  &  &  &  &  & $ $ \\  
\hhline{|-||*{14}{-}}
\vspace{1mm}
\includegraphics[scale=0.07]{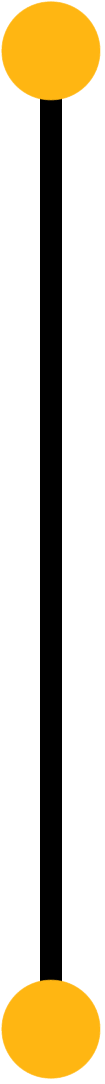}   & 2  & $-1$ &  &  &   &  &  &   &  &  &  &  &  & $ $  \\
\hhline{|-||*{14}{-}}
\vspace{1mm}
\includegraphics[scale=0.07]{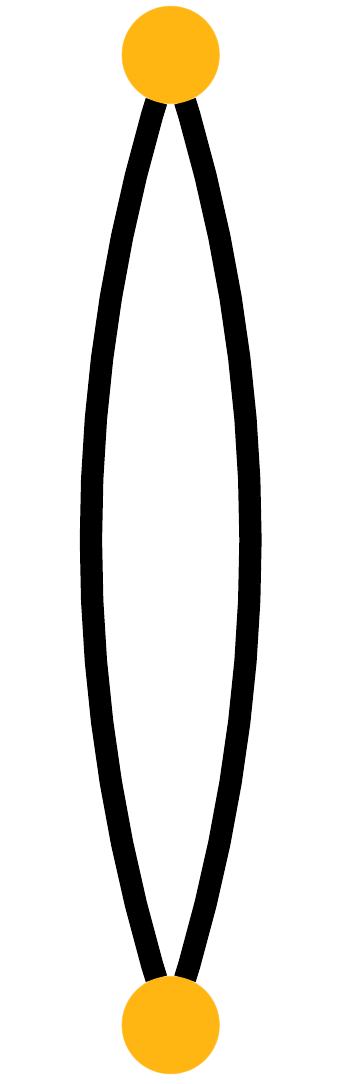}   & 4 & $-4$  & 1 & ~& ~ & ~ & ~& ~  & ~& ~& ~& ~ & ~& $ $ \\
\includegraphics[scale=0.07]{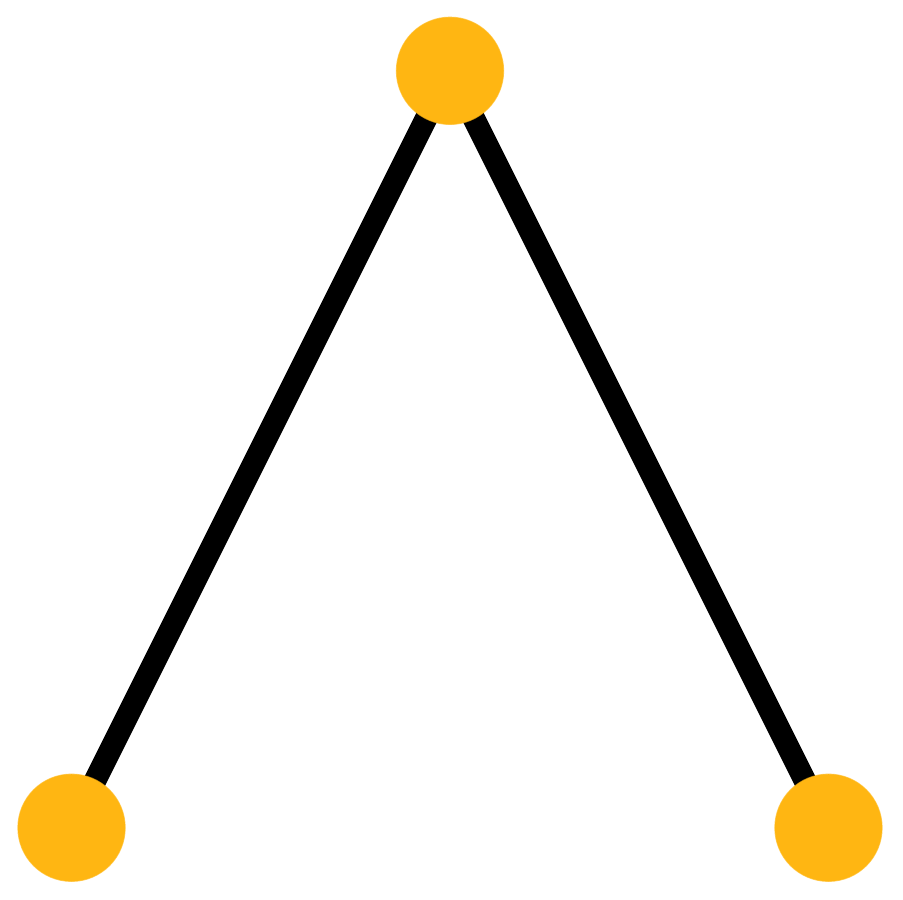}   &  4 & $-4$ & ~& 1 & ~ & ~& ~ & ~ & ~ & ~ & ~& ~& ~ & $ $ \\
\includegraphics[scale=0.07]{graphs/2_1_1_euclidean.pdf}
\includegraphics[scale=0.07]{graphs/2_1_1_euclidean.pdf}   & 4& $-4$& ~& ~& 1  & ~& ~& ~ & ~& ~ & ~ & ~ & ~ & $ $ \\
\hhline{|-||*{14}{-}}
\vspace{1mm}
\includegraphics[scale=0.07]{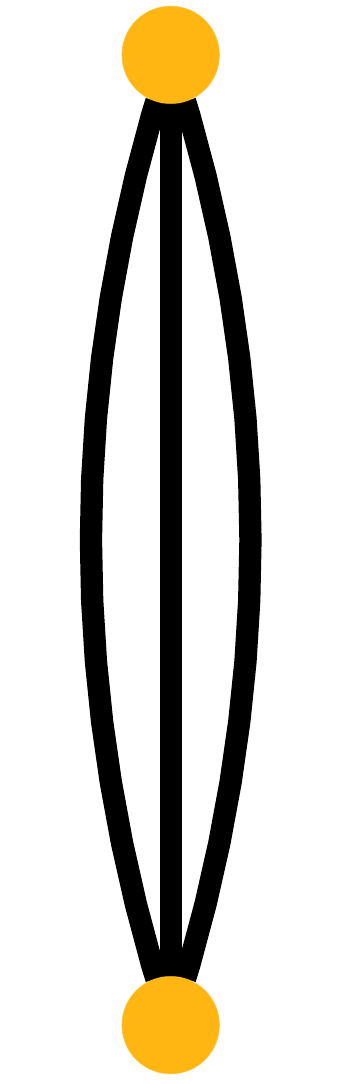}   & 8& $-12$& 6& ~& ~ & $-1$ & ~& ~ & ~& ~& ~& ~& ~& $ $ \\
\includegraphics[scale=0.07]{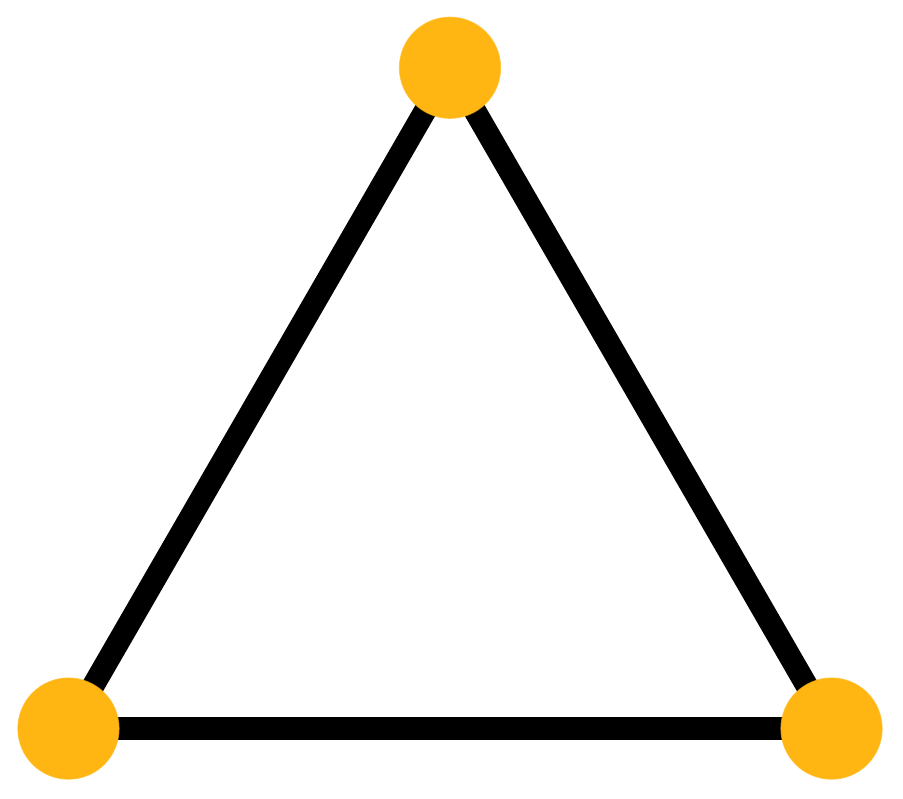}   & 8& $-12$& ~& 6& ~& ~& $-1$ & ~ & ~& ~& ~& ~& ~& $ $ \\
\includegraphics[scale=0.07]{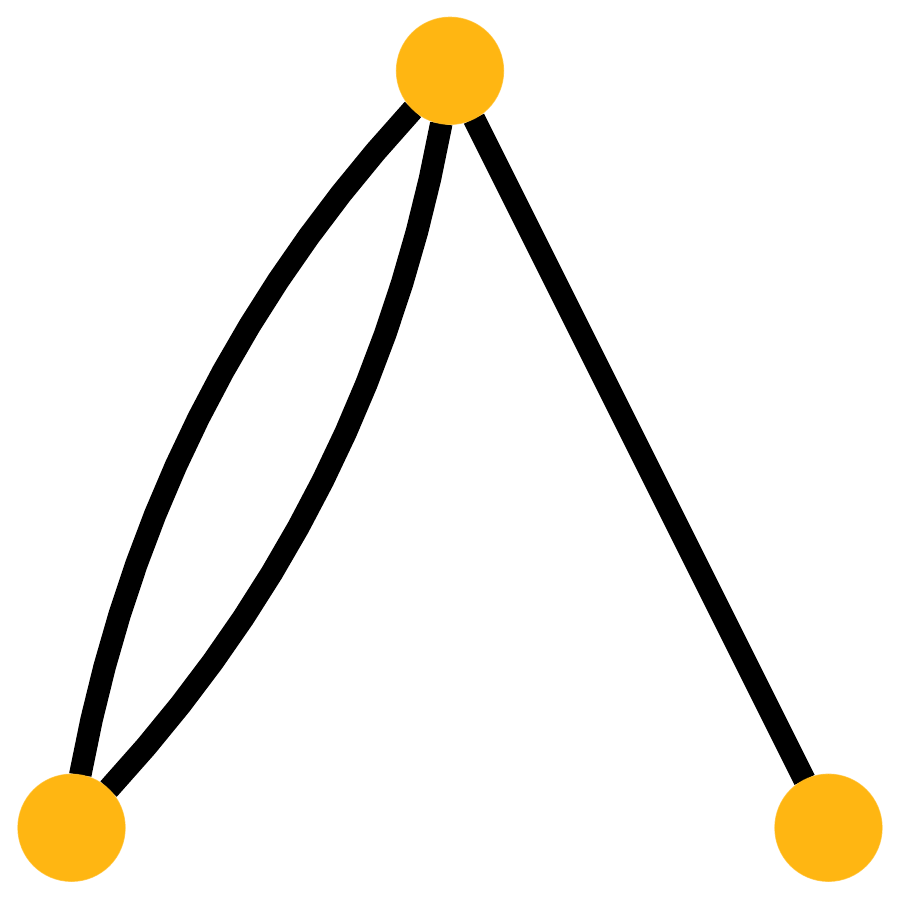}   & 8& $-12$& 2& 4 & ~ & ~& ~& $-1$  & ~& ~& ~& ~& ~& $ $ \\
\includegraphics[scale=0.09]{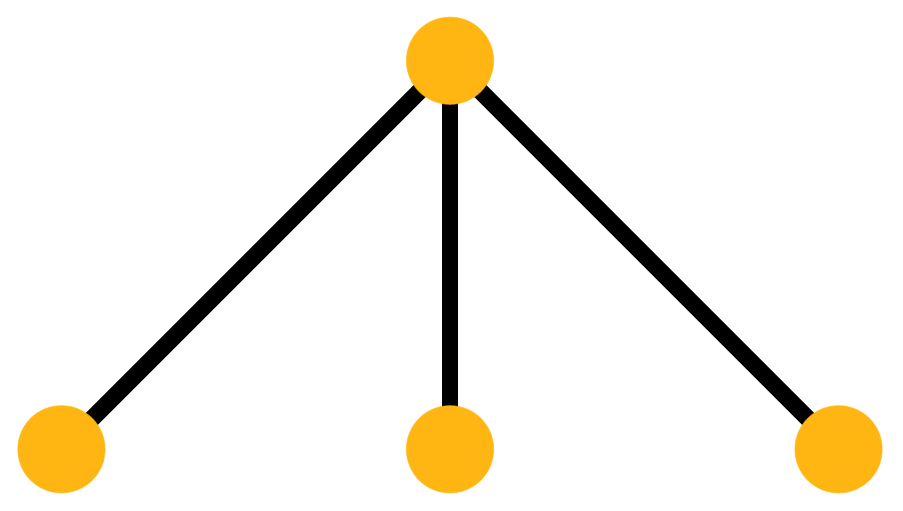}   & 8& $-12$ & ~& 6& ~ & ~& ~& ~ & $-1$ & ~& ~& ~& ~& $ $ \\
\includegraphics[scale=0.07]{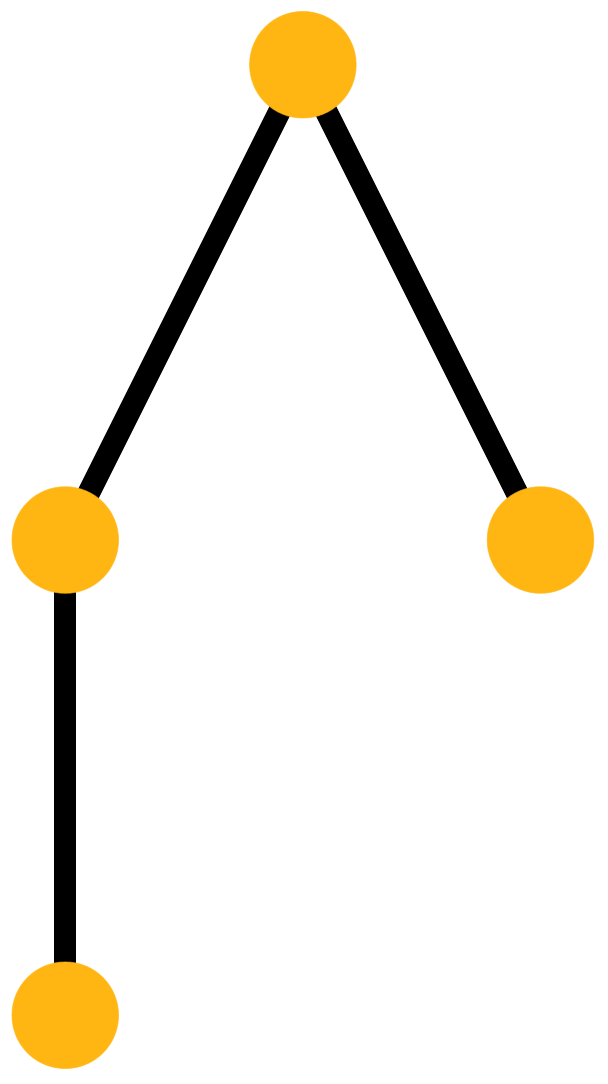}   & 8& $-12$ & ~& 4& 2 & ~& ~& ~ & ~& $-1$ & ~& ~& ~& $ $ \\
\includegraphics[scale=0.07]{graphs/2_1_1_euclidean.pdf}\includegraphics[scale=0.07]{graphs/2_2_1_euclidean.pdf}   & 8& $-12$& 2& ~& 4 & ~& ~& ~ & ~& ~& $-1$& ~ & ~& $ $ \\
\includegraphics[scale=0.07]{graphs/2_1_1_euclidean.pdf}\includegraphics[scale=0.07]{graphs/3_2_1_euclidean.pdf}  & 8& $-12$& ~& 2& 4 & ~& ~& ~ & ~& ~& ~& $-1$& ~ & $ $ \\
\includegraphics[scale=0.07]{graphs/2_1_1_euclidean.pdf} \includegraphics[scale=0.07]{graphs/2_1_1_euclidean.pdf} \includegraphics[scale=0.07]{graphs/2_1_1_euclidean.pdf}   & 8& $-12$& ~& ~& 6 & ~& ~& ~ & ~& ~& ~ & ~& $-1$& $ $ \\
\hhline{|-||*{14}{-}}
\vspace{1mm}
\includegraphics[scale=0.07]{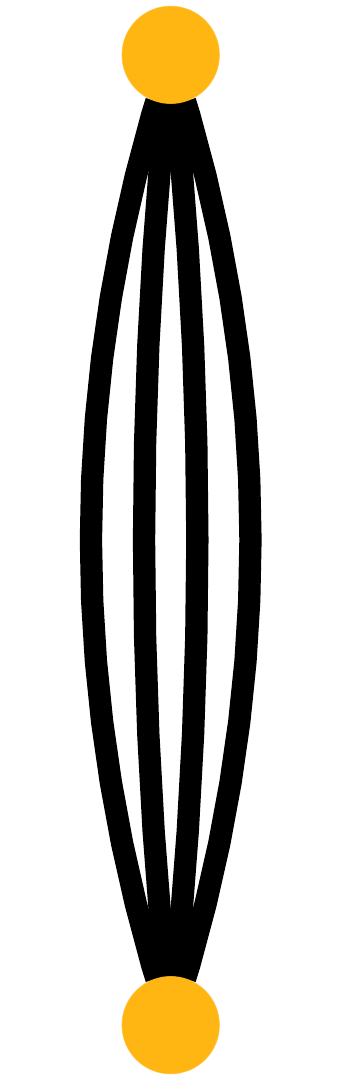}  & 16& $-32$& 24& ~& ~ & $-8$& ~& ~ & ~& ~& ~ & ~& ~& $1$ \\
\includegraphics[scale=0.07]{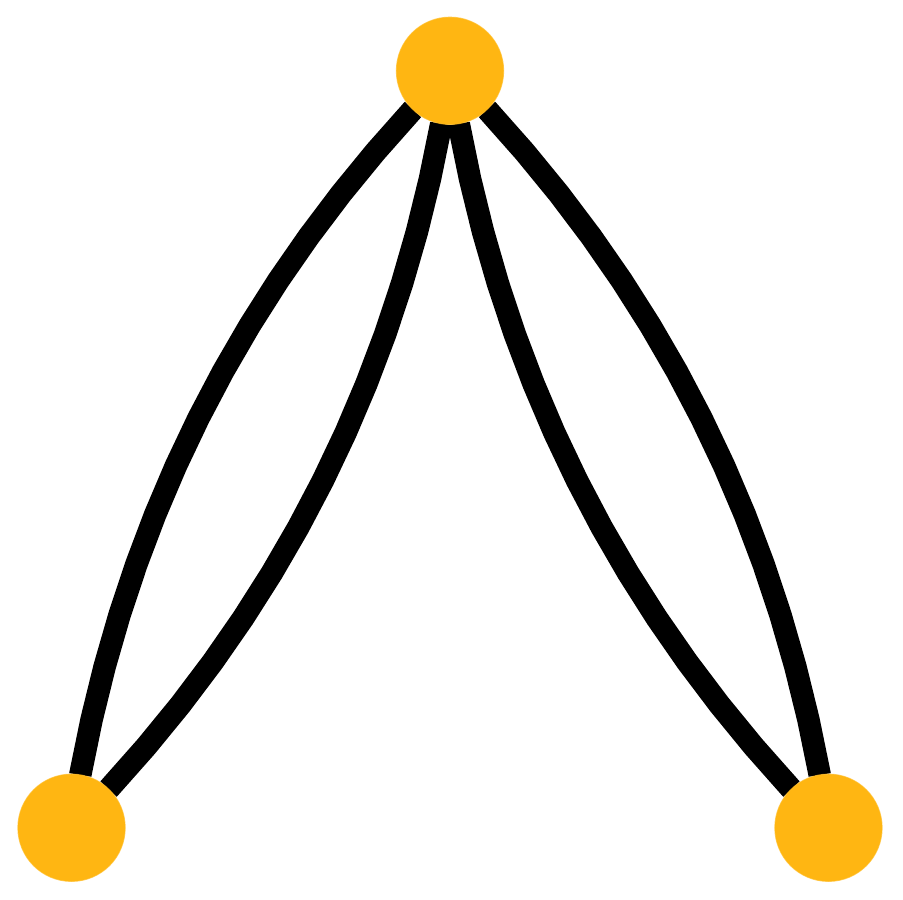}  & 16& $-32$& 8& 16& ~ & ~& ~& $-8$ & ~& ~& ~ & ~& ~& $ $ \\
\includegraphics[scale=0.07]{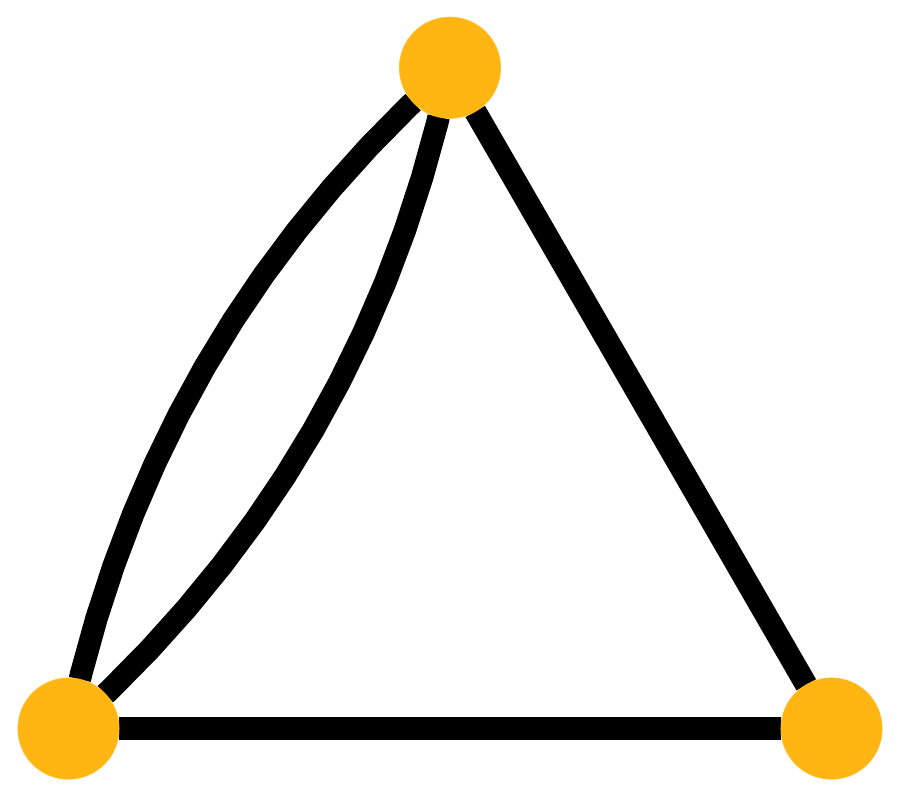}  & 16& $-32$& 4& 20& ~ & ~& $-4$& $-4$ & ~& ~& ~ & ~& ~& $ $ \\
\includegraphics[scale=0.07]{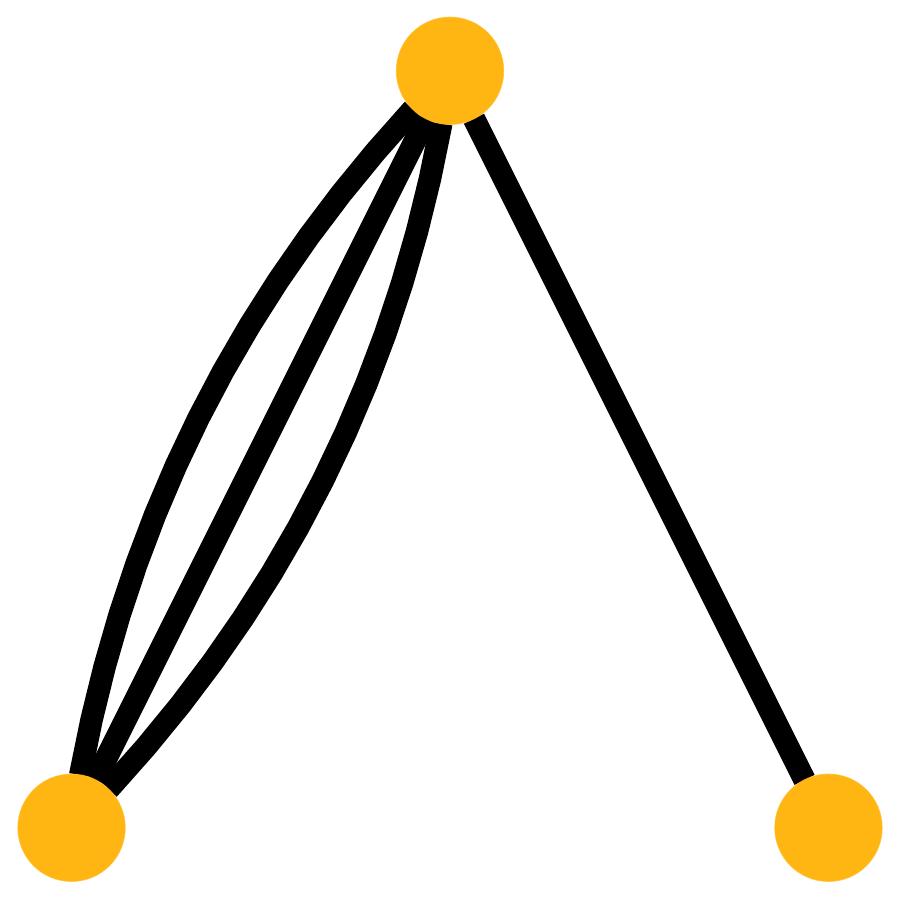}  & 16& $-32$& 12& 12& ~ & $-2$ & ~& $-6$& ~ & ~& ~ & ~& ~& $ $ \\
\includegraphics[scale=0.07]{graphs/4_4_1_euclidean.pdf}  & 16& $-32$& ~& 16& 8 & ~& ~& ~ & ~& $-8$& ~ & ~& ~& $ $ \\
\includegraphics[scale=0.07]{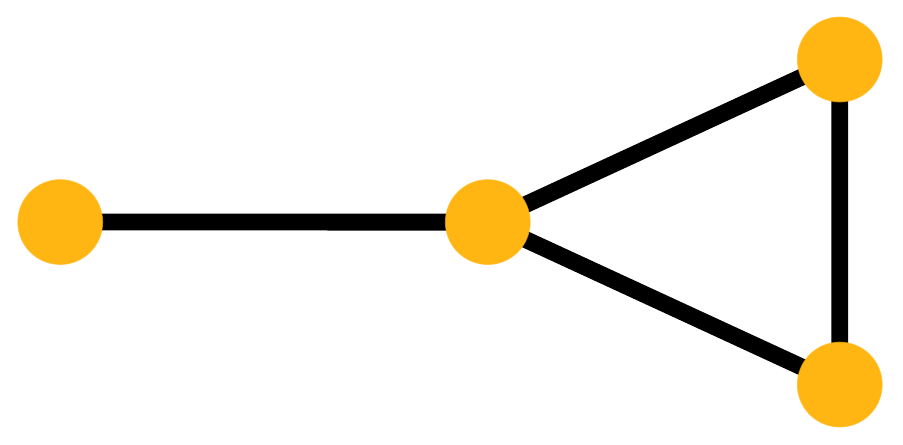}  & 16& $-32$& ~& 20& 4 & ~& $-2$& ~ & $-2$& $-4$ & ~ & ~& ~& $ $ \\
\includegraphics[scale=0.07]{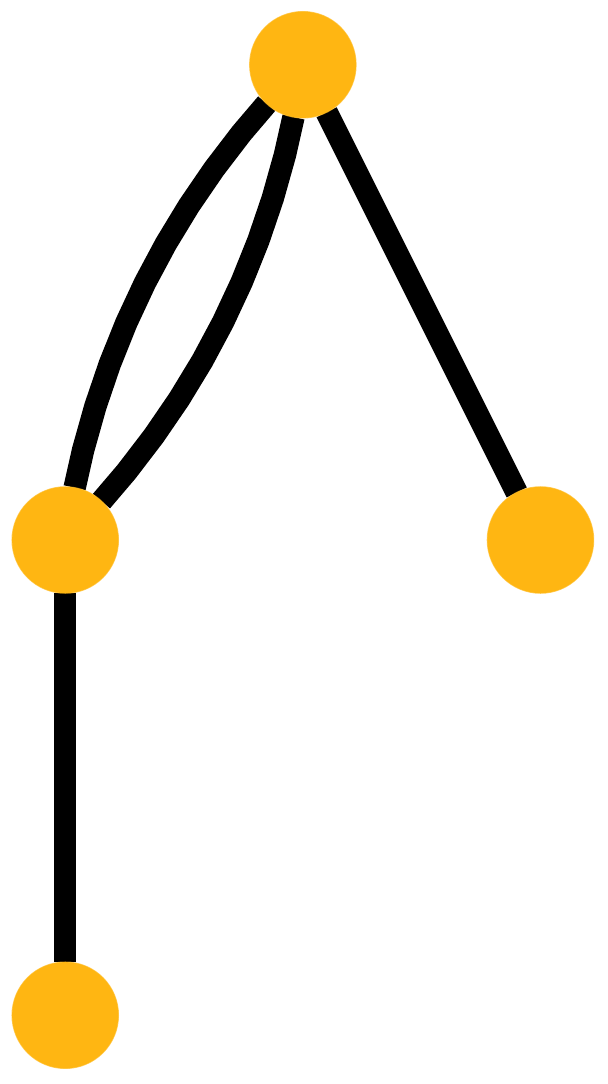}  & 16& $-32$& 4& 16& 4 & ~& ~& $-4$ & ~& $-4$& ~ & ~& ~& $ $ \\
\includegraphics[scale=0.07]{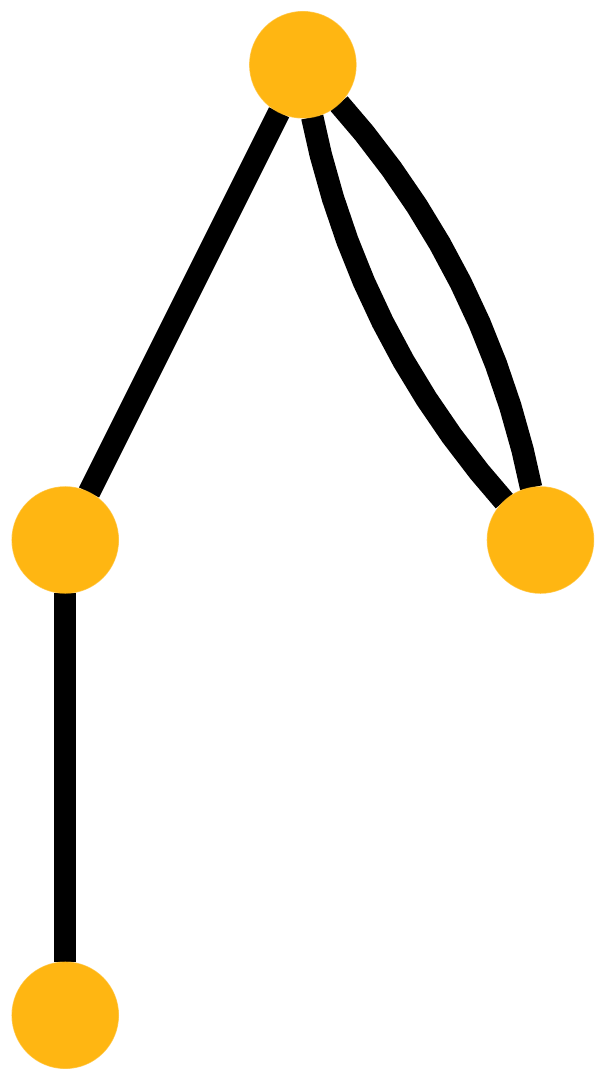}  & 16& $-32$& 4& 12& 8 & ~& ~& $-2$ & ~& $-4$ & $-2$ & ~& ~& $ $ \\
\includegraphics[scale=0.07]{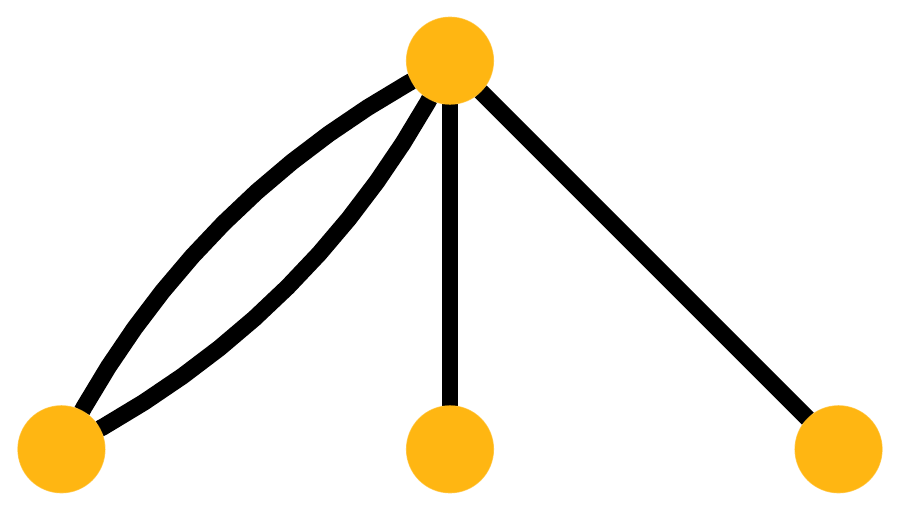}  & 16& $-32$& 4& 20& ~ & ~& ~& $-4$ & $-4$& ~& ~ & ~& ~& $ $ \\
\includegraphics[scale=0.07]{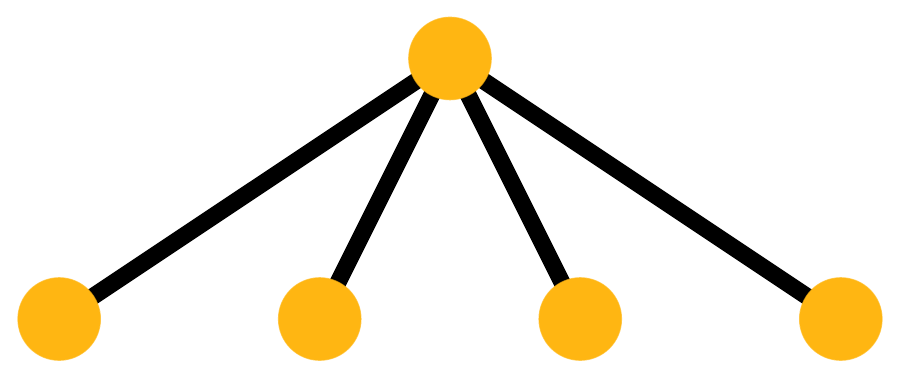}  & 16& $-32$& ~& 24& ~& ~& ~& ~ & $-8$& ~& ~ & ~& ~& $ $ \\
\includegraphics[scale=0.07]{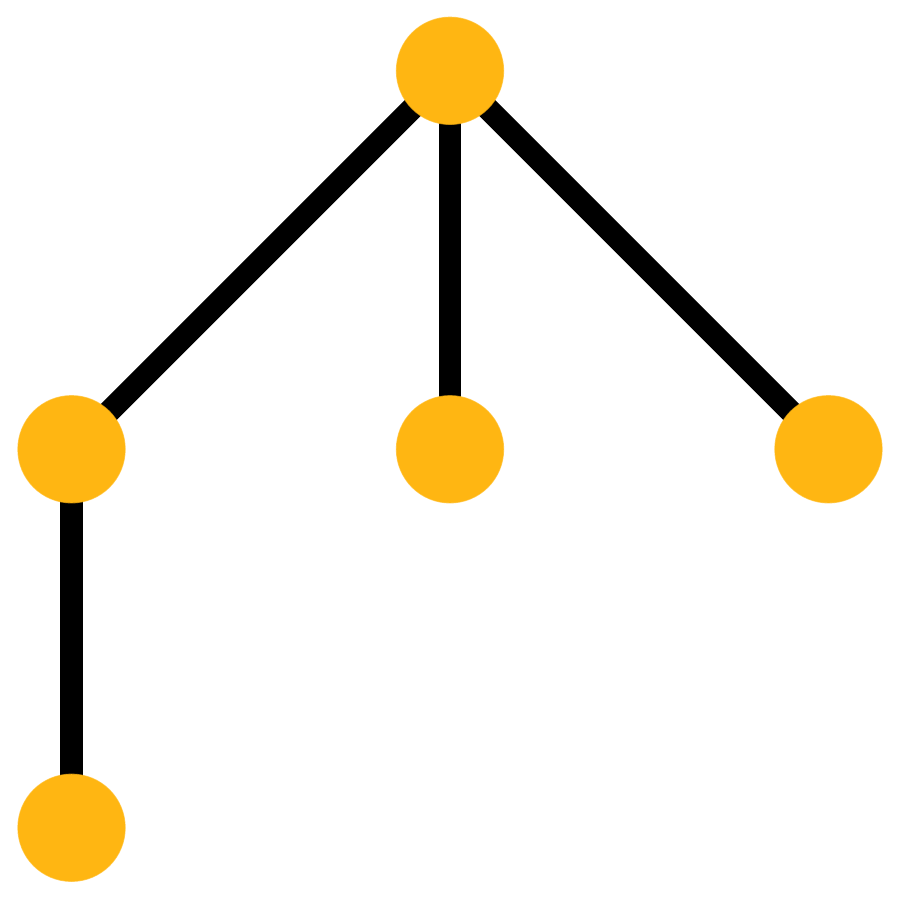}  & 16& $-32$& ~& 16& 8 & ~& ~& ~ & $-2$ & $-4$ & ~ & $-2$& $~$& $ $  \\
\includegraphics[scale=0.07]{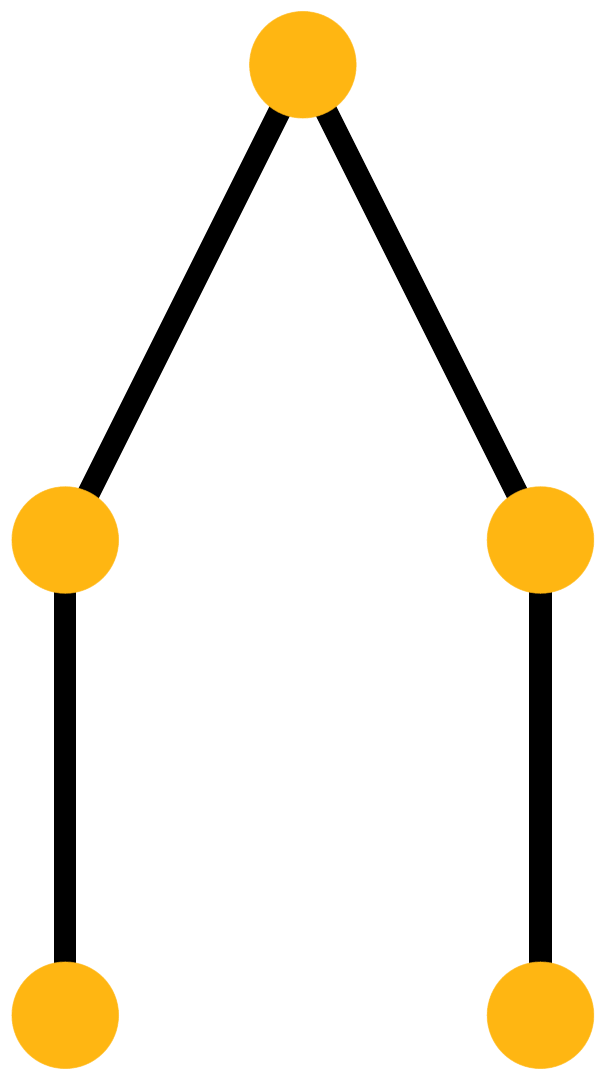}  & 16& $-32$& ~& 12& 12 & ~& ~& ~ & ~& $-4$ & ~ & $-4$& $~$& $ $ \\
$\vdots$ & $\vdots$ & $\vdots$ & $\vdots$ & $\vdots$ & $\vdots$ & $\vdots$ & $\vdots$ & $\vdots$ & $\vdots$ & $\vdots$ & $\vdots$ & $\vdots$ & $\vdots$ & $\ddots$
\end{tabular}
\caption{All contractions of generalized sphericity tensors (left column) as linear combinations of EFPs (top row).
The entries of the table are the specific linear coefficients of each EFP for a particular sphericity contraction.
This table can be used to translate between linear combinations of sphericity contractions and EFPs by matrix multiplication.
Shown are all graphs up to the connected graphs with five edges, though in principle it can be extended arbitrarily.
Note that a single diagonal of ones continues for the $d=4$ graphs and is not shown explicitly.}
\label{tab:graphmat}
\end{table*}

Next, we apply the $3\times3$ Cayley-Hamilton theorem from \Eq{eq:CH33} to $\Theta_2$.
Multiplying \Eq{eq:CH33} through by $\Theta_2$, we have:
\begin{equation}
\Theta_2^4 - \Theta_2^3 + \frac12(1 - \tr\Theta_2^2)\Theta_2^2 - \Theta_2\det \Theta_2 = 0.
\end{equation}
Substituting in for the $\det\Theta_2$ (from tracing over \Eq{eq:CH33}) and taking the trace of this expression gives rise to one identity.
Multiplying \Eq{eq:CH33} on both sides by $\Theta_1$ gives rise to a separate identity.
These two identities are:
\begin{align}\label{eq:sphid1}
0 &=6\, \begin{gathered}\includegraphics[scale=0.14]{graphs/4_4_1_euclidean.pdf}\end{gathered}
 - 16 \begin{gathered}\includegraphics[scale=0.14]{graphs/3_3_1_euclidean.pdf}\end{gathered}
- 3\begin{gathered}\includegraphics[scale=0.14]{graphs/2_2_1_euclidean.pdf}\end{gathered}\begin{gathered}\includegraphics[scale=0.14]{graphs/2_2_1_euclidean.pdf}\end{gathered}
+ 24 \begin{gathered}\includegraphics[scale=0.14]{graphs/2_2_1_euclidean.pdf}\end{gathered} - 16, \\
0 & = 6\, \begin{gathered}\includegraphics[scale=0.14]{graphs/5_4_3_euclidean.pdf}\end{gathered}
- 12\, \begin{gathered}\includegraphics[scale=0.14]{graphs/4_3_2_euclidean.pdf}\end{gathered}
- 3  \begin{gathered}\includegraphics[scale=0.14]{graphs/3_2_1_euclidean.pdf}\end{gathered}\begin{gathered}\includegraphics[scale=0.14]{graphs/2_2_1_euclidean.pdf}\end{gathered}
-2  \begin{gathered}\includegraphics[scale=0.14]{graphs/3_3_1_euclidean.pdf}\end{gathered}\begin{gathered}\includegraphics[scale=0.14]{graphs/2_1_1_euclidean.pdf}\end{gathered}
\nonumber\\ &\quad+ 12 \begin{gathered}\includegraphics[scale=0.14]{graphs/3_2_1_euclidean.pdf}\end{gathered}
+ 6 \begin{gathered}\includegraphics[scale=0.14]{graphs/2_2_1_euclidean.pdf}\end{gathered}\begin{gathered}\includegraphics[scale=0.14]{graphs/2_1_1_euclidean.pdf}\end{gathered}
- 8\, \begin{gathered} \includegraphics[scale=0.14]{graphs/2_1_1_euclidean.pdf}\end{gathered}.
\label{eq:sphid2}
\end{align}
In this analysis, we used a Cayley-Hamilton approach for simplicity.
In general, these graphical identities can be systematically derived from the tensor antisymmetrization identity in \Eq{eq:antisymid} by enumerating the possible antisymmetrizations (here, over four or more indices) of each graph structure.

To translate these identities among sphericity tensors  into identities among EFPs, we develop a look-up table which writes out each contraction of sphericity tensors as a specific linear combination of EFPs.
This can be worked out in the moment picture by recursively writing a Lorentzian contraction as a Euclidean contraction plus an additional term of opposite sign (with the appropriate factors of two).
Our results are summarized in \Tab{tab:graphmat} for all graphs up through the connected graphs with four edges.
With this table, we can quickly translate between the sphericity and energy flow pictures.
As an example, the identities in \Eqs{eq:sphid1}{eq:sphid2} give rise to the following identities among the EFPs:
\begin{align}
0 & = 6\, \begin{gathered}\includegraphics[scale=0.14]{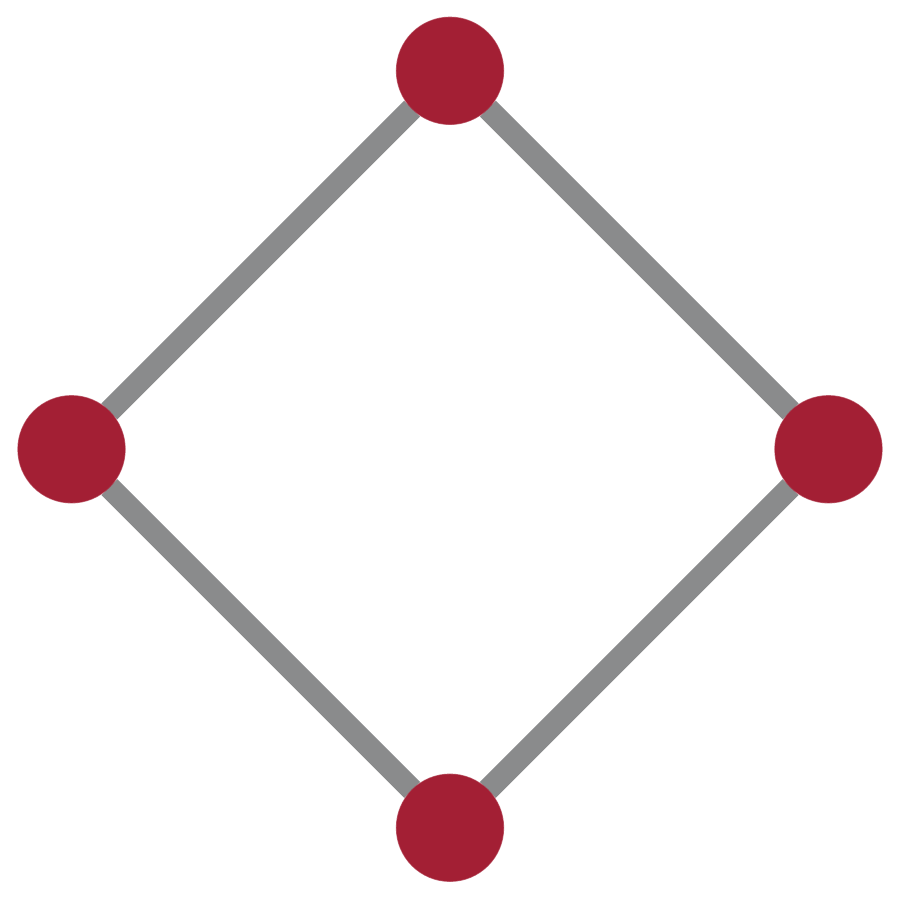}\end{gathered}
+ 16 \begin{gathered}\includegraphics[scale=0.14]{graphs/3_3_1.pdf}\end{gathered}
-3 \begin{gathered}\includegraphics[scale=0.14]{graphs/2_2_1.pdf}\end{gathered}\begin{gathered}\includegraphics[scale=0.14]{graphs/2_2_1.pdf}\end{gathered}
\nonumber\\ &\quad - 48\, \begin{gathered}\includegraphics[scale=0.14]{graphs/4_3_2.pdf}\end{gathered}
+ 24 \begin{gathered}\includegraphics[scale=0.14]{graphs/2_2_1.pdf}\end{gathered} \begin{gathered}\includegraphics[scale=0.14]{graphs/2_1_1.pdf}\end{gathered}
 ,\label{eq:efpid1}
 \end{align}
 \begin{align}
0 & =6\,  \begin{gathered}\includegraphics[scale=0.14]{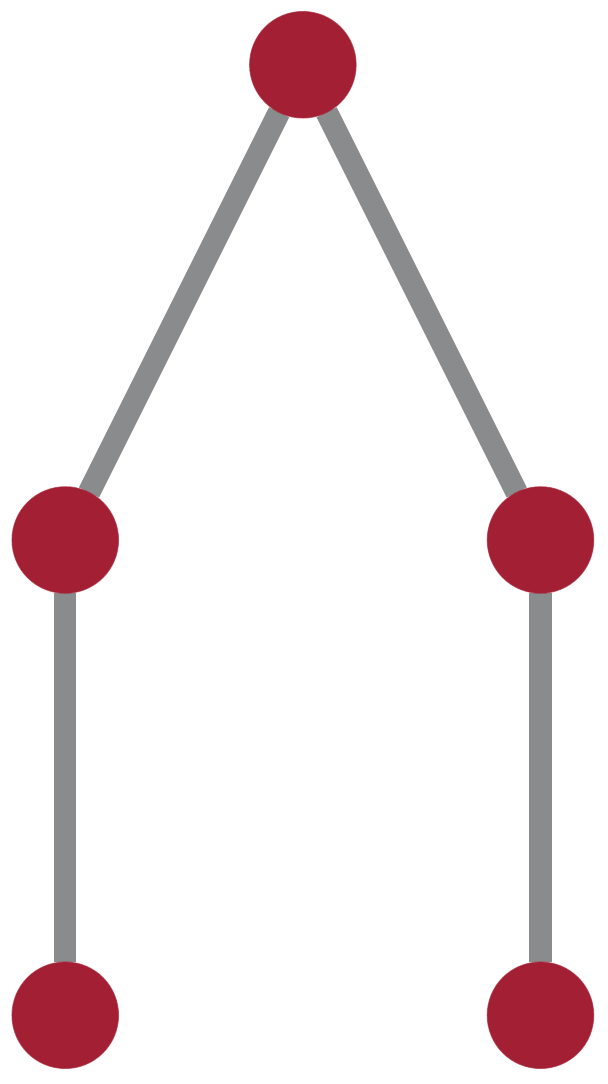}\end{gathered}
- 12\, \begin{gathered}\includegraphics[scale=0.14]{graphs/4_3_2.pdf}\end{gathered}
-3  \begin{gathered}\includegraphics[scale=0.14]{graphs/3_2_1.pdf}\end{gathered} \begin{gathered}\includegraphics[scale=0.14]{graphs/2_2_1.pdf}\end{gathered}
\nonumber\\ &\quad-2 \begin{gathered}\includegraphics[scale=0.14]{graphs/3_3_1.pdf}\end{gathered}\begin{gathered}\includegraphics[scale=0.14]{graphs/2_1_1.pdf}\end{gathered}
+ 4 \begin{gathered}\includegraphics[scale=0.14]{graphs/3_3_1.pdf}\end{gathered}
+ 6  \begin{gathered}\includegraphics[scale=0.14]{graphs/2_2_1.pdf}\end{gathered} \begin{gathered}\includegraphics[scale=0.14]{graphs/2_1_1.pdf}\end{gathered}
.\label{eq:efpid2}
\end{align}
Hence the finite dimension identities of the three-dimensional sphericity tensor induce related identities among the EFPs, with an interesting interplay between dimensionalities.


We can use similar reasoning to make contact with the $C$ and $D$ parameters.
From \Eq{eq:cddef}, these event shapes are related to the trace and determinant of the sphericity tensor via $C = \frac32((\tr\Theta)^2 - \tr\Theta^2)$ and $D = 27 \det \Theta$.
Using \Eq{eq:CH33} to express the determinant of $\Theta$, we can write these observables in terms of the sphericity graphs in the following way:
\begin{align}\label{eq:Ceuc}
C & = - \frac38\begin{gathered}\includegraphics[scale=0.14]{graphs/2_2_1_euclidean.pdf}\end{gathered} + \frac32,\\
D & = \frac{9}{8}\begin{gathered}\includegraphics[scale=0.14]{graphs/3_3_1_euclidean.pdf}\end{gathered} - \frac{27}{8}\begin{gathered}\includegraphics[scale=0.14]{graphs/2_2_1_euclidean.pdf}\end{gathered} + \frac{9}{2}.
\label{eq:Deuc}
\end{align}
We can use \Tab{tab:graphmat} to translate \Eqs{eq:Ceuc}{eq:Deuc} into the energy flow picture.
Doing so yields the EFP expressions for the $C$ and $D$ parameters already presented in \Eqs{eq:Clor}{eq:Dlor}.

\bibliography{efms}

\end{document}